\documentclass[journal,letterpaper,twocolumn,final,10pt]{IEEEtran}

\usepackage{amsthm}
\usepackage{amsmath}
\usepackage{amsfonts}
\usepackage{dsfont}
\usepackage{mathrsfs}
\usepackage{pifont}
\usepackage{amssymb}
\usepackage{verbatim}
\usepackage{upgreek}
\usepackage{color}
\usepackage[final]{graphicx}
\usepackage{algorithmic}
\usepackage{algorithm}
\usepackage{epsfig}
\usepackage{eucal}
\usepackage{cite}
\graphicspath{{Figures/}}

\newcommand{\bbm}{\begin{bmatrix}}
\newcommand{\ebm}{\end{bmatrix}}

\newcommand{\etr}[1]{{\mathrm{etr}}\left\{#1\right\}}

\newcommand{\bit}{\begin{itemize}}
\newcommand{\eit}{\end{itemize}}

\newcommand{\ben}{\begin{enumerate}}
\newcommand{\een}{\end{enumerate}}

\newcommand{\bdesc}{\begin{description}}
\newcommand{\edesc}{\end{description}}

\newcommand{\bea}{\begin{array}}
\newcommand{\eea}{\end{array}}

\newcommand{\tr}{\mbox{\rm Tr}\, }

\newcommand{\beqa}{\begin{eqnarray}}
\newcommand{\eeqa}{\end{eqnarray}}

\newcommand{\ds}{\displaystyle}

\newcommand{\Comment}[1]{}

\def\R{{\mathds R}}

\def\C{{\mathds C}}

\def\cC{\mbox{$\CMcal C$}}

\def\cH{\mbox{$\CMcal H$}}
\def\cL{\mbox{$\mathcal L$}}
\def\cN{\mbox{$\CMcal N$}}

\newcommand{\be}{\begin{equation}}
\newcommand{\ee}{\end{equation}}

\newcommand{\bzero}{{\mbox{\boldmath $0$}}}

\newcommand{\bai}{\bar{i}}

\newcommand{\boa}{{\mbox{\boldmath $a$}}}
\newcommand{\bob}{{\mbox{\boldmath $b$}}}

\newcommand{\bm}{{\mbox{\boldmath $m$}}}

\newcommand{\bor}{{\mbox{\boldmath $r$}}}

\newcommand{\bu}{{\mbox{\boldmath $u$}}}
\newcommand{\bv}{{\mbox{\boldmath $v$}}}
\newcommand{\bx}{{\mbox{\boldmath $x$}}}

\newcommand{\bq}{{\mbox{\boldmath $q$}}}
\newcommand{\bz}{{\mbox{\boldmath $z$}}}
\newcommand{\bgamma}{{\mbox{\boldmath $\gamma$}}}
\newcommand{\bA}{{\mbox{\boldmath $A$}}}

\newcommand{\bI}{{\mbox{\boldmath $I$}}}

\newcommand{\bM}{{\mbox{\boldmath $M$}}}

\newcommand{\bP}{{\mbox{\boldmath $P$}}}

\newcommand{\bR}{{\mbox{\boldmath $R$}}}
\newcommand{\bS}{{\mbox{\boldmath $S$}}}

\newcommand{\bU}{{\mbox{\boldmath $U$}}}

\newcommand{\bX}{{\mbox{\boldmath $X$}}}

\newcommand{\bZ}{{\mbox{\boldmath $Z$}}}

\newcommand{\dmax}{\begin{displaystyle}\max\end{displaystyle}}
\newcommand{\dmin}{\begin{displaystyle}\min\end{displaystyle}}

\newcommand{\test}{\mbox{$
\begin{array}{c}
\stackrel{ \stackrel{\textstyle \cH_{1}}{\textstyle >} }{
\stackrel{\textstyle <}{\textstyle \cH_{0}} }
\end{array}
$}}

\title{Adaptive Detection of Coherent Radar Targets in the Presence of Noise Jamming}

\author{Pia Addabbo,~\IEEEmembership{Member,~IEEE}, Olivier Besson,\\ Danilo Orlando,~\IEEEmembership{Senior Member,~IEEE},  and Giuseppe Ricci,~\IEEEmembership{Senior Member,~IEEE}
\thanks{Pia Addabbo is with Universit\`a degli Studi ``Giustino Fortunato", viale Raffale Delcogliano, 12, 82100 Benevento, Italy. E-mail: {\tt p.addabbo@unifortunato.eu} }
\thanks{Olivier Besson is with Institut Sup\'erieur de l'A\'eronautique et de l'Espace, University of Toulouse, Toulouse, France.  E-mail:  {\tt olivier.besson@isae-supaero.fr} }\\
\thanks{Danilo Orlando is with Universit\`a degli Studi ``Niccol\`o Cusano'', 
        Via Don Carlo Gnocchi, 3,  00166 Roma, Italy.
        E-mail: {\tt danilo.orlando@unicusano.it}}
\thanks{Giuseppe Ricci is with the Dipartimento di Ingegneria dell'Innovazione,
        Universit\`a del Salento, Via Monteroni, 73100 Lecce, Italy.
        E-mail: {\tt giuseppe.ricci@unisalento.it}.
        }
}

\begin{document}

\normalsize

\vspace{2cm}

\maketitle

\begin{abstract}
In this paper, we devise adaptive decision schemes to detect targets competing against clutter and {\em smart} noise-like jammers (NLJ)
which illuminate the radar system from the sidelobes. Specifically, the considered class of NLJs generates a pulse of 
noise (noise cover pulse) that is triggered by and concurrent with the received uncompressed pulse in order to mask 
the skin echo and, hence, to hide the true target range. The detection problem is formulated as a binary hypothesis test and
two different models for the NLJ are considered. Then, ad hoc modifications of the generalized likelihood ratio test are exploited
where the unknown parameters are estimated by means of cyclic optimization procedures. The performance analysis is carried out using
simulated data and proves the effectiveness of the proposed approach for both situations where the NLJ is either active or switched off.
\end{abstract}

\begin{IEEEkeywords}
Adaptive radar detection, alternating estimation, generalized likelihood ratio test, electronic countermeasure, electronic counter-countermeasures, 
noise cover pulse, noise-like jammers.
\end{IEEEkeywords}


\section{Introduction}
\label{sec:Introduction}
Electronic countermeasures (ECMs) are active techniques aimed at protecting
a platform from being detected and tracked by the radar \cite{ScheerMelvin}.
This is accomplished through two approaches: masking and deception.
Noncoherent jammers or noise-like jammers (NLJs) attempt to mask targets generating nondeceptive interference 
which blends into the thermal noise of the radar receiver. As a consequence, the radar sensitivity is degraded due to the increase of the
constant false alarm rate threshold which adapts to the higher level of noise \cite{antennaBased,ScheerMelvin}. In addition,
this increase makes more difficult to discover that jamming is taking place \cite{EW101,Farina-Handbook}.
On the other hand, the coherent jammers (CJs) transmit low-duty cycle signals intended to inject false information into the radar processor.
Specifically, they are capable of receiving, modifying, amplifying, and retransmitting the radar's own signal to create 
false targets maintaining radar's range, Doppler, and angle far away from the true position of the platform under 
protection \cite{Farina-Handbook,giniGrecoDRFM,antennaBased,ScheerMelvin}.

Nowadays, radar designers have developed defense strategies referred to as electronic counter-countermeasures (ECCMs)
which are aimed at countering the effects of the enemy's ECM and eventually succeeding in the intended mission. Such techniques
can be categorized as antenna-related, transmitter-related, receiver-related, and signal-processing-related depending on the main radar subsystem where
they take place \cite{Farina-Handbook}. The reader is referred to \cite[and references therein]{Farina-Handbook} for a detailed description of the major
ECCM techniques.

The first line of defense against jamming is represented by the radar antenna, whose beampattern can be suitably exploited and/or shaped to eliminate
sidelobe false targets or to attenuate the power of NLJs entering from the antenna sidelobes. In this context, 
famous antenna-related techniques capable of  preventing  jamming signals from entering through the radar sidelobes are the so-called
sidelobe blanking (SLB) and sidelobe canceling (SLC) \cite{DeVito-Farina-Sanzullo-Timmoneri}.
In particular, suppression of NLJs can be accomplished via an SLC system.
SLC uses an array of auxiliary antennas to adaptively estimate the direction of arrival  and the power of the jammers
and, subsequently, to modify the receiving pattern of the radar antenna placing nulls
in the jammers' directions. SLB and SLC can be jointly used to face with NLJs and CJs contemporaneously impinging on the sidelobes 
of the victim radar \cite{Farina-Timmoneri-Tosini}. In \cite{DeVito-Farina-Sanzullo-Timmoneri} it is also shown that
a data dependent threshold, based on \cite{Kalson}, outperforms a cascade of SLC and SLB stages.
The detector proposed in \cite{Kalson} is a special case of the more general class of tunable (possibly space-time) detectors
which have been shown to be an effective means to attack detection of mainlobe targets or rejection of CJs notwithstanding the presence 
of NLJs and clutter \cite{Kalson}-\!\!\!\cite{BOR-Morgan}. As a matter of fact, such solutions can be viewed as signal-processing-related ECCMs.
A way to design tunable receivers relies on the so-called two-stage architecture;
such schemes are formed by cascading two detectors (usually with opposite behaviors in terms of selectivity):
the overall one declares the presence of a target in the cell under test only when data
survive both detection thresholdings  \cite{Richmond-Asilomar98}-\!\!\cite{Richmond-2}, \cite{DeMaioRAO}-\!\!\cite{BOR-Morgan}.
Such detectors can also be used as classifiers: in this case, the first stage is less selective than the second one and it
is used to discriminate between the null hypothesis and the alternative that a structured signal is present.
In case of detection, the second stage is aimed at discrimination between
mainlobe and sidelobe signals, as explicitly
shown in \cite{Gini-Farina-Greco} for the adaptive sidelobe blanker (ASB).
Adaptive detection and discrimination between useful signals and
CJs in the presence of thermal noise, clutter, and possible
NLJ has also been addressed in \cite{BFOR}. Therein the CJ is assumed to belong to the orthogonal complement of the space spanned by the nominal steering vector
(after whitening by the true covariance matrix of the composite disturbance). This approach, based on a modified adaptive beamformer orthogonal rejection test (ABORT),
see also \cite{BOR-Morgan,W-ABORT}, allows to investigate the discrimination capabilities of  adaptive arrays
when the CJ is not necessarily confined to the ``sidelobe beam pattern,'' but might also be a mainlobe deception jammer.
A network of radars can be exploited to combat ECM signals.
In this case, it is reasonable that, for a given CUT,
only a subset of the radars receives ECM signals (CJs) as considered in \cite{CR}.

\begin{figure}[H]
\begin{center}
\includegraphics[scale=0.36]{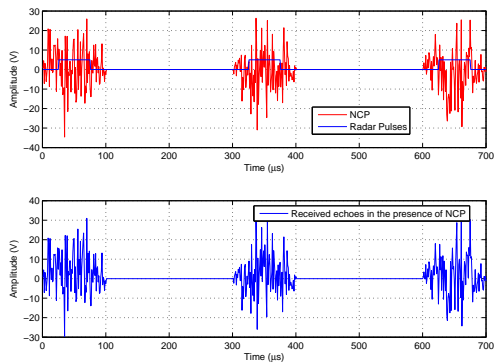}
\caption{Operating principle of NCP.}
\label{fig:NCP}
\end{center}
\end{figure}

Herein, we address adaptive detection in presence of noise cover pulse (NCP) jamming.
The NCP is an ECM technique belonging to the class of noise-like jamming.
Specifically, this kind of ECM generates a pulse of noise that is triggered by and concurrent with the received uncompressed pulse (see Figure \ref{fig:NCP}).
To this end, several received radar pulses are used to estimate the pulse width (PW) and the pulse repetition interval (PRI) to predict the arrival time instant of the next pulse of the victim radar. The transmitted noise power is strong enough to mask the skin echo even after the radar performs 
the pulse compression, which is used to enhance the range resolution. It follows that, since the length of the transmitted pulse is 
much higher than the duration of a range bin, the NCP creates an extended-range return spread over many range bins that hides the true target range.
Thus, it becomes of vital importance for a radar system to counteract the effects of an NCP attack. An ECCM technique against NCP is represented
by the cover pulse channel (CPC) \cite{van1982applied}, which consists in using an auxiliary physical channel to track the NCP transmission rather
than the skin return from the target. The main drawbacks of this technique are the degradation of the high-range resolution associated with the narrow pulses which result from the compression process and the exploitation of additional hardware resources. In order to overcome such limitations, 
in this paper we devise a signal-processing-related ECCM capable of detecting targets which compete against a NCP, while satisfying the 
original system requirements on range resolution. Besides, the proposed solution by its nature can 
reside in the signal processing unit of the system without the need of additional hardware. From a mathematical point of view, we formulate the detection
problem as a binary hypothesis test where primary data (namely those containing target returns) are formed 
by a set of range bins which is representative of the uncompressed pulse length and such that target return is located in only 
one bin whereas all the primary range bins are contaminated by the NCP. As for the NCP, we consider two models. In the first case, the NCP is represented  
as a rank-one modification of the interference covariance matrix (ICM), while in the second case the presence of the NCP is accounted for by including
a deterministic structured component in all the range bins. Moreover, we assume that a set of training samples are available to estimate 
the clutter and noise components of the ICM. These data are collected using a suitable number of guard cells surrounding those under test and related
to the uncompressed pulse length.
Then, we derive adaptive architectures exploiting {\em ad hoc} modifications of the generalized likelihood ratio test (GLRT) 
design criterion where the unknown parameters are estimated resorting to an alternating procedure.
Specifically, we leverage the cyclic optimization paradigm described in \cite{Stoica_alternating}.
Finally, we present numerical examples which highlight the effectiveness of the proposed solutions also in 
comparison with existing architectures which are somehow compatible with the considered problem.

The remainder of the paper is organized as follows: next section is
devoted to the problem formulation and to the description of the
two different models for the NCP. Section \ref{Sec:Designs} contains the derivation of the detection architectures, whereas
Section \ref{Sec:Assessment} provides the performance assessment of the detectors (also in comparison to natural competitors).
Concluding remarks and future research tracks are given in Section \ref{Sec:Conclusions}. 

\subsection{Notation}
In the sequel, vectors and matrices are denoted by boldface lower-case and upper-case letters, respectively.
The symbols $\det(\cdot)$, $\tr(\cdot)$, $\etr \cdot$, $(\cdot)^*$, $(\cdot)^T$, $(\cdot)^\dag$ denote the determinant, trace, 
exponential of the trace, complex conjugate, transpose, and conjugate transpose, respectively. As to numerical sets, $\R$ is the set of real numbers, $\R^{N\times M}$ is the Euclidean space of $(N\times M)$-dimensional real matrices (or vectors if $M=1$), $\C$ is the set of complex numbers, and $\C^{N\times M}$ is the Euclidean space of $(N\times M)$-dimensional complex matrices (or vectors if $M=1$). The symbols $\Re\left\{ z \right\}$ and $\Im\left\{ z \right\}$ indicate the real and imaginary parts of the complex number $z$, respectively. $\bI_N$ stands for the $N \times N$ identity matrix, while $\bzero$ is the null vector or matrix of proper dimensions. 
Let $f(\bx)\in\R^{N\times 1}$ be a scalar-valued function of vector argument, then $\partial f(\bx)/\partial \bx$ denotes the gradient of $f(\cdot)$ with respect to $\bx$ arranged in a column vector.
The Euclidean norm of a vector is denoted by $\|\cdot\|$. 
The $(k,l)$-entry (or $l$-entry) of a 
generic matrix $\bA$ (or vector $\boa$) is denoted by $\bA(k,l)$ (or $\boa(l)$). 
The acronym IID means independent and identically 
distributed while the symbol $E[\cdot]$ denotes statistical expectation.
Finally, we write $\bx\sim\cC\cN_N(\bm, \bM)$ if $\bx$ is an $N$-dimensional complex normal vector with mean $\bm$ and positive definite covariance matrix $\bM$.

\section{Problem Statement}
\label{sec_prob}

Assume that the radar is equipped with a linear array formed by $N$ antennas to sense the environment. For each sensor, the incoming signal 
is downconverted to baseband and, then, convolved with a conjugate time-reversed copy of the transmitted waveform (matched filter). 
The output of this filter is sampled to form the range bins of the area under surveillance. Thus, each range bin is represented by 
an $N$-dimensional complex vector. 
In what follows, we assume that the signal received from the cell under test (CUT) can be interference only, i.e.,  
thermal noise, clutter, and a possible NCP jamming, or a noisy version of the signal backscattered by
a coherent target.

As stated in Section \ref{sec:Introduction}, the NCP is an ECM technique belonging to the class of noise-like jamming.
Commonly,  on the radar side, the action of the NCP jammer leads to an increase of the noise level over many range bins hiding the true
target range.

\begin{figure}[H]
\begin{center}
\includegraphics[scale=0.36]{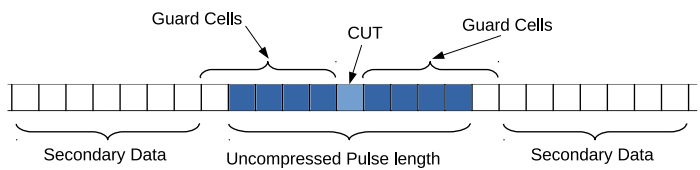}
\caption{Range bins partitioning.}
\label{fig:DataCollection}
\end{center}
\end{figure}

In order to model this situation, we denote by  $\bar{i}$ the integer indexing the CUT and by $\Omega=\{\bar{i}-H_1,\ldots,\bar{i}+H_2\}$ a set of integers indexing the range bins
contaminated by the NCP jammer which also include the CUT. The number of range bins after and before the CUT that are contaminated by the NCP jammer is not necessarily the same due to possible uncertainty in the PW and PRI estimates.
Moreover, such parameters are not known at the radar receiver, but an educated guess is possible.  Thus, in the following we do not address the problem of determining $H_1$ and $H_2$, but assume that $H_1$ and $H_2$ and, hence,  $H=H_1+H_2+1$, the number of contaminated cells, is known.
Additionally,
we assume that a set of $K\geq N$ secondary data, representative of thermal noise plus clutter only, is collected by the 
radar using a number of guard cells reflecting the length of the uncompressed pulse (see Figure \ref{fig:DataCollection}).


With the above model in mind, denote by $\bz_{\bai}\in\C^{N\times 1}$, $\bz_i\in\C^{N\times 1}$ with $i\in\Omega\setminus\{\bai\}$, and $\bor_k\in\C^{N\times 1}$
with $k=1,\ldots,K$, the vector containing the returns from the CUT, the vectors contaminated by NCP jammer, but free of target components, and 
the secondary data, respectively. For further developments we assume that such vectors are statistically independent.
Then, the problem of detecting the possible presence of a coherent return from a given cell 
is formulated in terms of the following hypothesis test
\be
\label{eqn:HypProb}
\left\{
\begin{array}{l}
\cH_0 : 
\left\{
\begin{array}{ll}
\bz_{\bai} \sim \cC\cN_N\left(\bzero, \bM+\bq \bq^{\dagger} \right), &   \\
\bz_{i} \sim \cC\cN_N\left(\bzero, \bM+\bq \bq^{\dagger} \right), \quad  i\in\Omega\setminus\{\bai\}, \\
\bor_{k} \sim \cC\cN_N\left(\bzero,  \bM \right), \quad  \quad \quad k=1, \ldots, K,  \\
\end{array}
\right.
\\
\\
\cH_1 : 
\left\{
\begin{array}{ll}
\bz_{\bai} \sim \cC\cN_N\left(\alpha \bv(\theta_T), \bM+\bq \bq^{\dagger} \right), &   \\
\bz_{i} \sim \cC\cN_N\left(\bzero, \bM+\bq \bq^{\dagger} \right), \quad  i\in\Omega\setminus\{\bai\}, \\
\bor_{k} \sim \cC\cN_N\left(\bzero,  \bM \right),  \quad \quad \quad k=1, \ldots, K,  \\
\end{array}
\right.
\end{array}
\right.
\ee
where
\begin{itemize}
\item
$\alpha \in \C$ is an unknown deterministic factor accounting for target response and channel effects;
\item
$\bv(\theta_T) =\frac{1}{\sqrt{N}}\left[1 \ e^{j\pi \sin(\theta_T)} \ \ldots \ e^{j\pi (N-1) \sin(\theta_T)} \right]^T\in \C^{N\times 1}$ is the 
known steering vector of the target with $\theta_T$ the angle of arrival of the target\footnote{Note that the steering corresponds 
to a uniform linear array with half-wavelength spacing.}; 
in the following, for brevity, we omit the dependence of $\bv$ 
on $\theta_T$.
\item
$\bM \in \C^{N \times N}$ is the unknown positive definite covariance matrix of thermal noise plus clutter;
\item
$\bq  \in \C^{N\times 1}$ is an unknown vector representing the contribution to the
noise covariance matrix of the NCP jamming.
\end{itemize}
Some definitions that will be used in the next developments for problem \eqref{eqn:HypProb} are now in order.
Let $\bZ_{\Omega,{\bai}}= [\bz_{\bai-H_1} \cdots \bz_{\bai-1} \ \bz_{\bai+1} \cdots \bz_{\bai+H_2}]$, 
$\bZ_{\alpha,\bai} = \left[ \bz_{\alpha,\bai} \ \bZ_{\Omega,\bai} \right]$ with $\bz_{\alpha,\bai}=\bz_{\bai}-\alpha \bv$, 
and $\bZ_{\bai} = \left[ \bz_{\bai} \ \bZ_{\Omega,\bai} \right]$.
Then, the probability density functions (PDFs) of $\bZ_{\bai}$ under $\cH_0$ and $\cH_1$ are given by
\be
\begin{array}{ll}
f_0(\bZ_{\bai} ; \bq, \bM)  \\ 
= \frac{1}{\left[\pi^N \det \left(\bM+ \bq \bq^{\dagger}\right) \right]^{H}} 
\etr{\displaystyle{
- \left(\bM+ \bq \bq^{\dagger}\right)^{-1} 
\bZ_{\bai} \bZ_{\bai}^{\dagger} 
}}
\end{array}
\ee
and
\be
\begin{array}{ll}
f_1(\bZ_{\bai} ; \alpha, \bq, \bM)  \\
= \frac{1}{\left[\pi^N \det \left(\bM+ \bq \bq^{\dagger}\right) \right]^{H}} 
\etr{\displaystyle{
- \left(\bM+ \bq \bq^{\dagger}\right)^{-1} 
\bZ_{\alpha,\bai} \bZ_{\alpha,\bai}^{\dagger} }},
\end{array}
\ee
respectively, whereas the PDF of $\bR=[\bor_1 \cdots \bor_K]$ under both hypotheses has the following expression
\be
f(\bR; \bM) = \frac{1}{\left[\pi^N \det \left(\bM\right) \right]^{K}} 
\etr{\displaystyle{
-\bM^{-1} 
\bR \bR^{\dagger} 
}}.
\ee
Finally, let us define the likelihood function of the unknown parameters under $\cH_i$, $i=0,1$, as
\begin{align}
\cL_0(\bq,\bM)&=f_0(\bz_{\bai}, \bZ_{\Omega,\bai} ; \bq, \bM),
\\
\cL_1(\alpha,\bq,\bM)&=f_1(\bz_{\bai}, \bZ_{\Omega,\bai} ; \alpha, \bq, \bM).
\end{align}
Now, we formulate the detection problem from another perspective. Specifically, observe that the radar system, at each dwell, collects a realization of the NCP. Thus, it is reasonable to compare \eqref{eqn:HypProb} with another detection problem formulated as
\begin{equation}
\label{eqn:comp_Problem1}
\left\{
\begin{array}{l}
\begin{aligned}
&\cH_{0}: \left\{
\begin{array}{l}
\bz_{\bai} 
\sim \cC\cN_N\left(\beta \bq, \bM \right),\\
\bz_i \sim \cC\cN_N\left(\beta_i \bq, \bM \right), \quad \quad i \in\Omega\setminus\{\bai\}, \\
\bor_{k} \sim \cC\cN_N\left(\bzero, \bM \right),  \quad  \quad k=1,\ldots,K, \\
\end{array}
\right.\\
&\cH_{1}: \left\{
\begin{array}{l}
\bz_{\bai} 
\sim \cC\cN_N\left(\alpha \bv+\beta \bq, \bM \right),\\
\bz_i \sim \cC\cN_N\left(\beta_i \bq, \bM \right),  \quad \quad i \in\Omega\setminus\{\bai\}, \\
\bor_{k} \sim \cC\cN_N\left(\bzero, \bM \right),  \quad  \quad k=1,\ldots,K, \\
\end{array}
\right.
\end{aligned}
\end{array}
\right.
\end{equation}
where 
\begin{itemize}
\item 
$\beta \in \C$ and $\beta_i \in \C$ are unknown deterministic factors representative of the different jammer amplitudes;
\item
$\bq  \in \C^{N\times 1}$ is an unknown deterministic vector representing the contribution 
of the NCP jamming.
\end{itemize}
Again, we have that
\begin{itemize}
\item
$\alpha \in \C$ is an unknown deterministic factor accounting for target response and channel effects;
\item
$\bv \in \C^{N\times 1}$ is the known steering vector of the target;
\item
$\bM \in \C^{N \times N}$ is the unknown positive definite covariance matrix of thermal noise plus clutter.
\end{itemize}
Furthermore, in this case,
the PDF of $\bZ_{\bai}$, under $\cH_{l}$, $l=0,1$, exhibits the following expression
\begin{multline}
f(\bz_{\bai},\bZ_{\Omega,\bai} ; l\alpha, \beta,\beta_i,
i \in\Omega\setminus\{\bai\},
\bM,\bq) = \frac{1}{\left[\pi^N \det(\bM)\right]^H}\\  \exp \left\{  -\tr\Bigg[\bM^{-1}
\Bigg(  (\bz_{\bai}-l\alpha\bv-\beta\bq) (\bz_{\bai}-l\alpha\bv-\beta\bq)^\dag \Bigg. \right. 
\\
 + \Bigg. \Bigg. \big. \sum_{i \in\Omega\setminus\{\bai\}} (\bz_i-\beta_i\bq) (\bz_i-\beta_i\bq)^\dag   \Bigg) 
\Bigg] \Bigg\},
\label{eqn:pdf_z}
\end{multline}
while the likelihood functions under $\cH_l, l=0,1,$ are given by
\be
\begin{array}{ll}
  \cL_0(\beta,\beta_i,i \in\Omega\setminus\{\bai\},\bM,\bq) \\ =  f(\bz_{\bai},\bZ_{\Omega,\bai} ; 0,\beta,\beta_i,i \in\Omega\setminus\{\bai\},\bM,\bq), \\
    \cL_1(\alpha, \beta,\beta_i,i \in\Omega\setminus\{\bai\},\bM,\bq) \\  = f(\bz_{\bai}, \bZ_{\Omega,\bai} ; \alpha, \beta,\beta_i,i \in\Omega\setminus\{\bai\},\bM,\bq).
\end{array}
\ee

\section{Detector Designs}
\label{Sec:Designs}

In this section, we device adaptive decision schemes for problems \eqref{eqn:HypProb} and \eqref{eqn:comp_Problem1}. To this end, observe that
we cannot apply the Neyman-Pearson criterion since parameters
$\alpha$, $\beta$, $\beta_i$, $\bM$ and $\bq$ are not known. For this reason, we have to resort to ad hoc solutions.
In particular, we adopt the two-step GLRT-based design procedure: first we derive the GLRT for known 
$\bM$; then we obtain an adaptive detector replacing the unknown matrix $\bM$ with an estimate 
based on secondary data. 
Thus, the main problem to solve is 
to discriminate between the interference-only-hypothesis $\cH_0$ 
and the signal-plus-interference-hypothesis $\cH_1$
based on $\bz_{\bai}$ and $\bZ_{\Omega,\bai}$ only (for known $\bM$).

\subsection{An adaptive architecture for problem \eqref{eqn:HypProb}}\label{sec_glrt_detector}

The GLRT for known $\bM$ is given by
\be
\frac{\dmax_{\alpha, \bq} \cL_1( \alpha, \bq, \bM)}
{\dmax_{\bq} \cL_0(\bq, \bM)} \test \eta,
\ee
where $\eta$ is the threshold\footnote{Hereafter, $\eta$ denotes any modification of the original threshold.} to be set according to the desired value of the probability of false alarm ($P_{fa}$).

Maximization of the PDF under $\cH_0$ can be conducted using the following identities
\begin{eqnarray}
\nonumber
\det \left(\bM+ \bq \bq^{\dagger}\right) \!\!\!\!\!&=&\!\!\!\!\! \det \left(\bM \right) 
\det \left(\bI_N+ \bM^{-1/2} \bq \bq^{\dagger} \bM^{-1/2} \right)
\\ &=&\!\!\!\!\! \det \left(\bM \right) 
\left(1+  \bu^{\dagger} \bu \right),
\label{eq:det}
\end{eqnarray}
where $\bu=\bM^{-1/2} \bq$
and 
\begin{align}
&  \tr \!\! \left[ \left(\bM+ \bq \bq^{\dagger}\right)^{-1} \!\!\bZ_{\bai} \bZ_{\bai}^{\dagger} \right] \nonumber \\
& = \tr \!\!
\left[ \bM^{-1/2}  \left(\bI+ \bu \bu^{\dagger}\right)^{-1}  \bX_{\bai} \bZ_{\bai}^{\dagger} \right]
\nonumber \\ 
& =
\tr \!\!
\left[ \left(\bI+ \bu \bu^{\dagger}\right)^{-1}  \bX_{\bai} \bX_{\bai}^{\dagger} \right]
\nonumber \\ & =
\tr \!\! \left[ \bX_{\bai} \bX_{\bai}^{\dagger}  - \frac{\bu \bu^{\dagger}}{1+ \bu^{\dagger} \bu} \bX_{\bai} \bX_{\bai}^{\dagger} \right]
\label{eq:trace}
\end{align}
where 
the last equality in equation~(\ref{eq:trace}) is obtained using the matrix inversion lemma while
$\bX_{\bai}=\bM^{-1/2} \bZ_{\bai} =
\bM^{-1/2} \left[ \bz_{\bai} \ \bZ_{\Omega,\bai} \right] =
[\bx_{\bai} \ \bX_{\Omega,\bai}]$.
Maximization under the $\cH_1$ hypothesis is conducted using 
identity~(\ref{eq:det}), but replacing (\ref{eq:trace})
with
\[
\tr \left[ \left(\bM+ \bq \bq^{\dagger}\right)^{-1} \bZ_{\alpha,\bai} \bZ_{\alpha,\bai}^{\dagger} \right] 
\]
\[
=
\tr \left[ \bX_{\alpha,\bai} \bX_{\alpha,\bai}^{\dagger} - \frac{\bu \bu^{\dagger}}{1+ \bu^{\dagger} \bu} \bX_{\alpha,\bai} \bX_{\alpha,\bai}^{\dagger} \right]
\]
where $\bX_{\alpha,\bai}=\bM^{-1/2} \bZ_{\alpha,\bai} =[ \bx_{\alpha,\bai} \ \bX_{\Omega,\bai} ]$
and, in particular, $\bx_{\alpha,\bai}=\bM^{-1/2} \left( \bz_{\bai} - \alpha \bv \right)$.
It follows that the likelihood functions under $\cH_0$ and $\cH_1$ can be re-written as
\be
\begin{array}{ll}
\cL_0( \bq, \bM) = \frac{1}{\left[\pi^N \det \left(\bM \right) \left(1+  \bu^{\dagger} \bu \right) \right]^{H}} \\
\times \etr{\displaystyle{-
 \bX_{\bai} \bX_{\bai}^{\dagger}  + \frac{\bu \bu^{\dagger}}{1+ \bu^{\dagger} \bu} \bX_{\bai} \bX_{\bai}^{\dagger} }}
 \end{array}
\ee
and
\be\label{eqn:likeihoodH1}
\begin{array}{ll}
\cL_1( \alpha, \bq, \bM) = \frac{1}{\left[\pi^N \det \left(\bM \right) \left(1+  \bu^{\dagger} \bu \right) \right]^{H}}\\
\times \etr{\ds{-
\bX_{\alpha,\bai} \bX_{\alpha,\bai}^{\dagger}  + \frac{\bu \bu^{\dagger}}{1+ \bu^{\dagger} \bu} \bX_{\alpha,\bai} \bX_{\alpha,\bai}^{\dagger} }},
\end{array}
\ee
respectively.

Now, we focus on the maximization of the PDF under $\cH_0$. To this end, observe that $\bu$ can be represented as
$\bu=\sqrt{p} \bu_0$ with $p={\bu^{\dagger} \bu}=\|\bu\|^2>0$ and, hence, $\| \bu_0\|=1$. For 
future reference, we also define by ${\cal S}$ the $N$-sphere centered at the origin with unit radius; thus, condition 
$\bu_0^{\dagger} \bu_0=\|\bu_0\|^2=1$ is equivalent to $\bu_0 \in {\cal S}$.

It follows that
\begin{eqnarray*}
\max_{\bq} \cL_0( \bq, \bM) &=& 
\frac{1}{\left[\pi^N \det \left(\bM \right) \right]^{H}} 
\etr{\displaystyle{-
 \bX_{\bai} \bX_{\bai}^{\dagger}}}
\\ &\times&
\max_{\bu_0, p}
\frac{1}{\left(1+  p \right)^{H}} 
\etr{\displaystyle{ \frac{p \bu_0 \bu_0^{\dagger}}{1+ p} \bX_{\bai} \bX_{\bai}^{\dagger} }}.
\end{eqnarray*}
Thus, for known $\bu_0$, maximizing $\cL_0$ with respect to $p$ is tantamount to
maximizing
\be
g(p)=
\frac{1}{(1+p)^{H}} \exp\left\{
\displaystyle{\frac{p}{1+p}} \bu_0^{\dagger} \bX_{\bai} \bX_{\bai}^{\dagger} \bu_0\right\}
\ee
with respect to $p\geq0$.
It can be shown that
the maximum is attained at
\be
\widehat{p}=
\left\{
\begin{array}{ll}
 \ds\frac{\bu_0^{\dagger} \bX_{\bai} \bX_{\bai}^{\dagger} \bu_0}{H}-1, 
& \mbox{if } \ds\frac{\bu_0^{\dagger} \bX_{\bai} \bX_{\bai}^{\dagger} \bu_0}{H}>1, \\
0, & \mbox{otherwise,}
\end{array}
\right.
\ee
and is given by
\be
\begin{array}{ll}
\max_{p \geq 0} g(p)\\ \!\!\!\!=\!\!
\left\{
\begin{array}{ll}
\!\!\!\!\! \ds\left[ \frac{H}{\bu_0^{\dagger} \bX_{\bai} \bX_{\bai}^{\dagger} \bu_0} \right]^{H}
\!\!\!\!\!  \exp\{\bu_0^{\dagger} \bX_{\bai} \bX_{\bai}^{\dagger} \bu_0 -H\}, &
\!\!\!\!\!  \mbox{if } \!\! \ds \frac{\bu_0^{\dagger} \bX_{\bai} \bX_{\bai}^{\dagger} \bu_0}{H}\! >\! 1, \\ \\
1, & \mbox{otherwise}.
\end{array}
\right.
\end{array}
\ee
Now, we let
$$
h(x)= 
\left\{
\begin{array}{ll}
\ds\left[ \frac{H}{x} \right]^{H}
e^{x -H}, &
x \in (H, +\infty), \\ \\
1, & x \in [0, H],
\end{array}
\right.
$$
and observe that it is a strictly increasing function of $x$ over $[H, +\infty)$ (and constant over $[0, H]$).
It follows that
to maximize $\cL_0$ with respect to $\bu$ it is sufficient to plug the maximizer
of  
$\bu_0^{\dagger} \bX_{\bai} \bX_{\bai}^{\dagger} \bu_0$ with respect to $\bu_0$ into 
$\dmax_{p \geq 0} g(p)$.
Using the Rayleigh-Ritz theorem \cite{HornJohnson}, we
obtain
\be
\dmax_{\bu_0 \in {\cal S}} \bu_0^{\dagger} \bX_{\bai} \bX_{\bai}^{\dagger} \bu_0
= \lambda_{1} \left(\bX_{\bai} \bX_{\bai}^{\dagger} \right),
\ee
where $\lambda_{1} \left( \cdot \right)$ denotes the maximum eigenvalue 
of the matrix argument and a maximizer for $\bu_0$ is a normalized eigenvector of the matrix $\bX_{\bai} \bX_{\bai}^{\dagger}$ corresponding to $\lambda_{1} \left(\bX_{\bai} \bX_{\bai}^{\dagger} \right)$.
Thus, we can conclude that
\begin{align}
& \max_{\bq} \cL_0( \bq, \bM) =
\frac{1}{\left[\pi^N \det \left(\bM \right) \right]^{H}} 
\etr{\displaystyle{-
 \bX_{\bai} \bX_{\bai}^{\dagger}}}
\nonumber \\ 
&\!\!\! \times
\left\{
\begin{array}{ll}
\!\!\!\! \ds\left[  \frac{H}{\! \lambda_{1} \!\! \left(\bX_{\bai} \bX_{\bai}^{\dagger} \right)} \right]^{H} \!\!\!\!\!\!
\exp\left\{\lambda_{1} \!\! \left(\bX_{\bai} \bX_{\bai}^{\dagger} \right) \!\! -\! H \! \right\}\!, &
\!\!\!\! \mbox{if } \ds\frac{\lambda_{1} \left(\bX_{\bai} \bX_{\bai}^{\dagger} \right)}{H}\!>\!1, \\ \\
1, & \mbox{otherwise}.
\end{array}
\right.
\end{align}

As for the optimization problem under $\cH_1$, let us 
compute the logarithm 
of the likelihood function \eqref{eqn:likeihoodH1} neglecting the terms independent of $\alpha$ and $\bu$ to obtain 
\begin{align}
\label{onecolEq}
g(\alpha,p,\bu_0)
=&-H\log(1+\bu^\dag\bu)\nonumber\\
&-\tr\left[ \left(\bI_N-\frac{\bu\bu^\dag}{1+\bu^\dag\bu}\right)\bS_{\Omega,\bai} \right] \nonumber\\
&-\tr\left[ \left(\bI_N-\frac{\bu\bu^\dag}{1+\bu^\dag\bu}\right)\bS_{\alpha,\bai} \right] \nonumber
\\
=& -H\log(1+\bu^\dag\bu)+\frac{\bu^\dag \bS_{\Omega,\bai} \bu}{1+\bu^\dag\bu}  \nonumber\\
&+ \frac{\left|\bx_{\alpha,\bai}^\dag\bu\right|^2}{1+\bu^\dag\bu} - \bx_{\alpha,\bai}^\dag\bx_{\alpha,\bai}
-\tr\left[ \bS_{\Omega,\bai} \right]
\nonumber
\\
=&-H\log(1+p)+\frac{p}{1+p} \bu_0^\dag \bS_{\Omega,\bai} \bu_0 \nonumber \\
&+ \frac{p}{1+p}|\bx_{\alpha,\bai}^\dag\bu_0|^2 - \bx_{\alpha,\bai}^\dag\bx_{\alpha,\bai}
-\tr\left[ \bS_{\Omega,\bai} \right],
\end{align}
 where $\bS_{\Omega,\bai}=\bX_{\Omega,\bai} \bX_{\Omega,\bai}^\dag$ and $\bS_{\alpha,\bai}=\bx_{\alpha,\bai} \bx_{\alpha,\bai}^\dag$.

It follows that maximizing $\cL_1( \alpha, \bq, \bM)$ 
with respect to $\alpha$ and $\bq$
is tantamount to
\be
\dmax_{\alpha,p,\bu_0} g(\alpha,p,\bu_0).
\ee

However, this joint maximization with respect to $\alpha$, $p$, and $\bu_0$ is not an analytically tractable problem at least to the best of authors' knowledge.
For this reason, we resort to a suboptimum approach 
relying on alternating maximization \cite{Stoica_alternating}.
Specifically, let us assume that $\bu_0=\bu_0^{(n)}$ and $p=p^{(n)}$ are known, then it is not difficult to show that
\begin{align}\label{eqn:alpha_estimate}
\alpha^{(n)}&=\arg\dmax_{\alpha} g\left(\alpha,p^{(n)},\bu_0^{(n)}\right) \nonumber
\\
&=\arg\dmin_{\alpha} \bx_{\alpha,\bai}^\dag \left[\bI_N -\frac{p^{(n)}}{1+p^{(n)}}\bu_0^{(n)}(\bu_0^{(n)})^\dag \right]\bx_{\alpha,\bai} \nonumber
\\
&= \ds\frac{\bv_0^\dag\left[\bI_N-\ds\frac{p^{(n)}}{1+p^{(n)}}\bu_0^{(n)}(\bu_0^{(n)})^\dag\right]\bx_{\bai}}
{\bv_0^\dag\left[\bI_N-\ds\frac{p^{(n)}}{1+p^{(n)}}\bu_0^{(n)}(\bu_0^{(n)})^\dag\right]\bv_{0}},
\end{align}
where $\bv_0=\bM^{-1/2}\bv$. Now, let us exploit $\alpha^{(n)}$ to estimate $\bu_0$ and $p$, namely to solve the problem
\be
\dmax_{p,\bu_0\in {\cal S}} g\left(\alpha^{(n)},p,\bu_0\right).
\ee
To this end, following the same line of reasoning as for $\cH_0$, we obtain that
\begin{align}\label{eqn:q_estimate}
\begin{array}{ll}
\begin{bmatrix}
p^{(n+1)}
\\
\bu_0^{(n+1)}
\end{bmatrix}
&=\arg\dmax_{p,\bu_0\in {\cal S}}g\left(\alpha^{(n)},p,\bu_0\right)\\
&=
\begin{bmatrix}
\max\{\lambda_1(\bS_{\Omega,\bai} + \bx_{\alpha^{(n)},\bai}\bx_{\alpha^{(n)},\bai}^\dag)/H-1,0\}
\\
\bob_1
\end{bmatrix},
\end{array}
\end{align}
where we remember that $\lambda_1(\cdot)$ is the maximum eigenvalue of the matrix argument, $\bx_{\alpha^{(n)},\bai}$ is obtained 
replacing $\alpha$ with $\alpha^{(n)}$ in $\bx_{\alpha,\bai}$, and $\bob_1$ is a normalized eigenvector corresponding 
to $\lambda_1(\bS_{\Omega,\bai} + \bx_{\alpha^{(n)},\bai}\bx_{\alpha^{(n)},\bai}^\dag)$.

Iterating the above estimation procedure, we come up with the following nondecreasing sequence
\begin{align}\label{eqn:sequence}
& \cL_1( \alpha^{(0)}, \bq^{(0)}, \bM)\leq \cL_1( \alpha^{(0)}, \bq^{(1)}, \bM)\nonumber\\
& \leq
\cL_1( \alpha^{(1)}, \bq^{(1)}, \bM)\leq\ldots\leq \cL_1( \alpha^{(n)}, \bq^{(n)}, \bM),
\end{align}
where we start using
for $\bq^{(0)}$ a normalized steering vector from a sidelobe direction and
$\bq^{(i)}=p^{(i)} \bM^{1/2} \bu_0^{(i)}$, $i=1,\ldots,n$. Since
the likelihood under $\cH_1$ is bounded from the above with respect to $\alpha$, $p$, $\bu_0$,
the nondecreasing sequence \eqref{eqn:sequence} converges as $n$ diverges and, hence, a 
suitable stopping criterion can be defined.
For example, a possible strategy might consist in continuing the procedure until
\be
\|\bq^{(n)}-\bq^{(n-1)}\|<\epsilon_q \quad \mbox{and/or} \quad |\alpha^{(n)}-\alpha^{(n-1)}|<\epsilon_{\alpha}.
\ee
Another approach might be that the alternating procedure terminates when $n>N_{max}$ with $N_{max}$
the maximum allowable number of iterations. We will use the latter stopping criterion with $N_{max}$ chosen in the next section.

To prove that the likelihood is bounded from the above, we re-write $g$ as the sum of three (bounded above) functions, namely as
\begin{align}
g(\alpha,p,\bu_0)
=&-H\log(1+p)+\frac{p}{1+p} \bu_0^\dag \bS_{\Omega,\bai} \bu_0 \nonumber \\
&+\frac{p}{1+p}|\bx_{\alpha,\bai}^\dag\bu_0|^2 - \bx_{\alpha,\bai}^\dag\bx_{\alpha,\bai} \nonumber
\\ 
=&g_1(p)+g_2(p,\bu_0)+g_3(\alpha,p,\bu_0)
\end{align}
with 
\begin{align}
g_1(p)&=-H\log(1+p) -\tr\left[ \bS_{\Omega,\bai} \right],
\\
g_2(p,\bu_0)&=\frac{p}{1+p} \bu_0^\dag \bS_{\Omega,\bai} \bu_0,
\\
g_3(\alpha,p,\bu_0)&=\frac{p}{1+p}|\bx_{\alpha,\bai}^\dag\bu_0|^2 - \bx_{\alpha,\bai}^\dag\bx_{\alpha,\bai}.
\end{align}
Then, it is sufficient to observe that
\begin{itemize}
\item
$g_1(p)\leq 0$, $\forall p \geq 0$;
\item 
$\forall p \geq 0, \bu_0 \in {\cal S}$ the second term $g_2(p,\bu_0)$ can be trivially upperbounded as
\begin{multline}
g_2(p,\bu_0)=\frac{p}{1+p} \bu_0^\dag \bS_{\Omega,\bai} \bu_0 
\\
\leq \bu_0^\dag \bS_{\Omega,\bai} \bu_0 (\leq \lambda_1(\bS_{\Omega,\bai}))
\end{multline}
and the right-most hand side attains a maximum since it is a continuous function of $\bu_0$
over a compact set ($\bu_0$ belongs to the $N$-sphere with unit radius);
\item 
the third term can be re-written as
\[
g_3(\alpha,p,\bu_0)=-\bx_{\alpha,\bai}^\dag \left(\bI_N-  \frac{p}{1+p}\bu_0 \bu_0^\dag\right)\bx_{\alpha,\bai};
\]
since the matrix $\bI_N-  \frac{p}{1+p}\bu_0 \bu_0^\dag$ is positive definite $\forall p \geq 0$ and 
$\bu_0 \in {\cal S}$,
it follows that $g_3(\alpha,p,\bu_0) \leq 0$, $\forall \alpha \in \C,p \geq 0,\bu_0 \in {\cal S}$.
\end{itemize}
Finally, $\bM$ can be estimated using secondary data $\bR$ as
\be
\widehat{\bM}=\frac{1}{K}\bR\bR^\dag.
\label{eq_estM}
\ee
This decision scheme is referred to in the following as random NCP detector (R-NCP-D).

\subsection{An adaptive architecture for problem \eqref{eqn:comp_Problem1}}
\label{sec_mod_glrt_detector}

In this case, the GLRT for known $\bM$ is given by
\be
\frac{\dmax_{\alpha, \beta,\beta_i, i \in\Omega\setminus\{\bai\},\bq}\cL_1(\alpha, \beta,\beta_i, i \in\Omega\setminus\{\bai\},\bM,\bq)}{\dmax_{\beta,\beta_i, i \in\Omega\setminus\{\bai\},\bq}\cL_0( \beta,\beta_i, i \in\Omega\setminus\{\bai\},\bM,\bq)}\test \eta.
\label{eq_glrtbis}
\ee

Focusing on the maximization of the PDF under $\cH_0$, we observe that maximizing $\cL_0$ is tantamount to maximizing
\begin{align}
& w(\beta,\beta_i, i \in\Omega\setminus\{\bai\}, \bq) = \nonumber \\ 
& \!\!\!\!
\!\!\etr{\!\!\displaystyle{
\! - \bM^{-1} \!\!
\Bigg[  (\bz_{\bai}-\beta\bq) (\bz_{\bai}-\beta\bq)^\dag  \! + \!\!\!\!\!   \sum_{i \in\Omega\setminus\{\bai\}} \!\!\!\! (\bz_i-\beta_i\bq) (\bz_i-\beta_i\bq)^\dag \! \Bigg]}\!\!}
\label{eq_w}
\end{align}
Furthermore, it can be proved that
\begin{align}
& \dmin_{\beta} (\bz_{\bai}-\beta\bq)^\dag \bM^{-1} (\bz_{\bai}-\beta\bq)  \nonumber\\
& = \bz_{\bai}^{\dag} \bM^{-1} \bz_{\bai} - \frac{\bz_{\bai}^{\dag} \bM^{-1} \bq \bq^{\dag} \bM^{-1} \bz_{\bai}}{\bq^{\dag}\bM^{-1}\bq}
\label{eq_minbeta}
\end{align}
and
\begin{align}
& \dmin_{\beta_i}\sum_{i \in\Omega\setminus\{\bai\}} (\bz_i-\beta_i\bq)^\dag \bM^{-1} (\bz_i-\beta_i\bq) \nonumber \\
& = \sum_{i \in\Omega\setminus\{\bai\}} \left( \bz_i^{\dag} \bM^{-1} \bz_i - \frac{\bz_i^{\dag} \bM^{-1} \bq \bq^{\dag} \bM^{-1} \bz_i}{\bq^{\dag}\bM^{-1}\bq} \right).
\label{eq_minbetai}
\end{align}
Thus, the maximization of $w$ with respect to $\beta$ and $\beta_i$ leads to  
\begin{align}
& \dmax_{\beta,\beta_i, i \in\Omega\setminus\{\bai\},\bq} w(\beta,\beta_i, i \in\Omega\setminus\{\bai\}, \bq)  
\nonumber \\
 =& \dmax_{\bq} \exp\left[
    \frac{\bz_{\bai}^{\dag} \bM^{-1} \bq \bq^{\dag} \bM^{-1} \bz_{\bai}}{\bq^{\dag}\bM^{-1}\bq} - \bz_{\bai}^{\dag} \bM^{-1} \bz_{\bai} \right. \nonumber\\
   & \left. + \sum_{i \in\Omega\setminus\{\bai\}} \left( \frac{\bz_i^{\dag} \bM^{-1} \bq \bq^{\dag} \bM^{-1} \bz_i}{\bq^{\dag}\bM^{-1}\bq} - \bz_i^{\dag} \bM^{-1} \bz_i  \right) 
   \right] 
\nonumber \\
 = &\dmax_{\bq} \exp \left\{  \frac{\bq^{\dag} \bM^{-1}\left[\bz_{\bai}\bz_{\bai}^{\dag}+
\sum_{i \in\Omega\setminus\{\bai\}}\bz_{i}\bz_{i}^{\dag}\right]\bM^{-1}\bq}{\bq^{\dag}\bM^{-1}\bq} \right. \nonumber \\
& \left. - \left( \bz_{\bai}^{\dag} \bM^{-1} \bz_{\bai} + \sum_{i \in\Omega\setminus\{\bai\}} \bz_i^{\dag} \bM^{-1} \bz_i \right) \right\} \nonumber \\
 = &\dmax_{\bu} \exp \left\{  \frac{\bu^{\dag} \left[\bx_{\bai}\bx_{\bai}^{\dag}+
\sum_{i \in\Omega\setminus\{\bai\}}\bx_{i}\bx_{i}^{\dag}\right]\bu}{\bu^{\dag}\bu} \right. \nonumber\\
&\left. - \left( \bx_{\bai}^{\dag}  \bx_{\bai} + \sum_{i \in\Omega\setminus\{\bai\}} \bx_i^{\dag}  \bx_i \right) \right\} \nonumber \\
 =&  \exp \left\{  \lambda_1 (\bS_1) - \tr (\bS_1) \right\}
\label{eq_maxq}
\end{align}
 where  $\bS_1=\bx_{\bai}\bx_{\bai}^{\dag}+\sum_{i \in\Omega\setminus\{\bai\}}\bx_{i}\bx_{i}^{\dag}$ 
and we used the Rayleigh-Ritz theorem \cite{HornJohnson}.
We also recall that $\bx_i=\bM^{-1/2}\bz_i$ and 
$\bu=\bM^{-1/2}\bq$.

Thus, we conclude that the compressed likelihood under  $\cH_0$
is given by
\begin{align}
& \dmax_{\beta,\beta_i, i \in\Omega\setminus\{\bai\},\bq}
\cL_0(\beta,\beta_i,i \in\Omega\setminus\{\bai\},\bM,\bq)\nonumber\\
& =
\frac{1}{\left[\pi^N \det(\bM)\right]^H}
\exp \left\{ \lambda_1 \left( \bS_1 \right) - \tr (\bS_1)  \right\}.
 \label{eq_maxcl0}
\end{align}
As far the maximization of the likelihood function under $\cH_1$ is concerned, we first focus on $\alpha$ and observe that
\be
\begin{aligned}
&\dmin_{\alpha} \left( (\bz_{\bai}-\alpha\bv-\beta\bq)^\dag  \bM^{-1} (\bz_{\bai}-\alpha\bv-\beta\bq) \right) \\
&= (\bz_{\bai} -\beta \bq )^{\dag} \bM^{-1} ( \bz_{\bai} -\beta \bq ) \\ & - \frac{(\bz_{\bai} -\beta \bq )^{\dag}\bM^{-1}\bv\bv^{\dag}\bM^{-1}(\bz_{\bai}-\beta \bq)}{\bv^{\dag}\bM^{-1}\bv}.
\end{aligned}
\label{eq_maxalpha}
\ee
Thus, it turns out that
\begin{align}
& \dmax_{\alpha}
\cL_1(\alpha,\beta,\beta_i,i \in\Omega\setminus\{\bai\},\bM,\bq)
=
\frac{1}{\left[\pi^N \det(\bM)\right]^H} \nonumber\\
& \times
\exp \Bigg\{ \frac{(\bz_{\bai} -\beta \bq )^{\dag}\bM^{-1}\bv\bv^{\dag}\bM^{-1}(\bz_{\bai}-\beta \bq)}{\bv^{\dag}\bM^{-1}\bv} \Bigg. \nonumber\\
& \left. -(\bz_{\bai} -\beta \bq )^{\dag} \bM^{-1} ( \bz_{\bai} -\beta \bq ) \right. \nonumber \\
& \left. -  \sum_{i \in\Omega\setminus\{\bai\}}(\bz_i-\beta_i\bq)^\dag \bM^{-1} (\bz_i-\beta_i\bq)\right\}.
\label{eq_h1}
\end{align}
To the best of authors' knowledge maximization of \eqref{eq_h1} with respect to the remaining parameters cannot be conducted in closed form. 
For this reason, we exploit another alternating optimization procedure. 
To this end, we first re-write the partially-compressed likelihood as
\begin{align}
& \dmax_{\alpha}
\cL_1(\alpha,\beta,\beta_i,i \in\Omega\setminus\{\bai\},\bM,\bq) \nonumber \\
& =
\frac{1}{\left[\pi^N \det(\bM)\right]^H} 
\exp \left[- h\left( \bu, \beta, \beta_i, i \in\Omega\setminus\{\bai\} \right) \right]
\label{eq:compressed_L1_alpha}
\end{align}
with
\begin{align}
& h\left( \bu, \beta, \beta_i, i \in\Omega\setminus\{\bai\} \right) \nonumber \\
 &= \left(\bx_{\bai}-\beta\bu\right)^{\dag} \bP_{\bv_0}^{\perp} \left(\bx_{\bai}-\beta\bu\right)  + \!\!\!\!\! \sum_{i \in\Omega\setminus\{\bai\}} \!\!\! \left(\bx_i-\beta_i\bu\right)^\dag  \left(\bx_i-\beta_i\bu\right)
\end{align}
where 
\be
\bP_{v_0}^{\perp} = \bI - \frac{\bv_0\bv_0^\dag}{\bv_0^\dag \bv_0},
\ee
and, for the reader ease, we recall that $\bu=\bM ^{-1/2}\bq$, $\bx_i=\bM^{-1/2}\bz_i$, $\bx_{\bai}=\bM^{-1/2}\bz_{\bai}$, and $\bv_0=\bM^{-1/2}\bv$.

Then, assuming that $\beta=\beta^{(n)}$ and $\beta_i=\beta_i^{(n)}$ are given, we can 
focus on the maximization with respect to $\bq$. 
To this end, setting to zero the first derivative\footnote{We make use of the following definition for the derivative of a 
real function $f(\alpha)$ with respect to the complex argument $\alpha=\alpha_r+j \alpha_i$,
$\alpha_r, \alpha_i \in \R$, \cite{VanTrees4} 
\be
\frac{\partial f(\alpha)}{\partial \alpha} = \frac{1}{2} \left[ \frac{\partial f(\alpha)}{\partial \alpha_r} 
- j \frac{\partial f(\alpha)}{\partial \alpha_i}\right].
\ee} 
of 
$$h\left( \bu, \beta^{(n)}, \beta_i^{(n)}, i \in\Omega\setminus\{\bai\} \right)$$  
with respect to $\bu$ leads to
\begin{align}
& -{\beta^{(n)}}^*\bP_{\bv_0}^{\perp}\bx_{\bai} +
\left|\beta^{(n)}\right|^2 \bP_{\bv_0}^{\perp} \bu \nonumber\\
& - \sum_{i \in\Omega\setminus\{\bai\}}{\beta_i^{(n)}}^* \bx_i + \sum_{i \in\Omega\setminus\{\bai\}} 
\left|\beta_i^{(n)}\right|^2 \bu  = 0;
\label{eq_derq0}
\end{align}
then
\begin{align}
\bu^{(n)} = & \arg\dmin_{\bu}\left\{ h\left( \bu, \beta^{(n)}, \beta_i^{(n)}, i \in\Omega\setminus\{\bai\} \right) \right\} \nonumber \\
 =& \left(\left|\beta^{(n)}\right|^2 \bP_{\bv_0}^{\perp}
+\sum_{i \in\Omega\setminus\{\bai\}}\left|\beta_i^{(n)}\right|^2\bI \right)^{-1} \nonumber \\
& \left({\beta^{(n)}}^*\bP_{\bv_0}^{\perp}\bx_{\bai}+\sum_{i \in\Omega\setminus\{\bai\}}{\beta_i^{(n)}}^*\bx_i\right).
\label{eq_minq0}
\end{align}

On the other hand, we can estimate $\beta^{(n+1)}$ and $\beta_i^{(n+1)}$, given $\bu^{(n)}$, exploiting a standard least-squares result, i.e.,
\begin{align}
\begin{bmatrix}
\beta^{(n+1)}
\\
\beta_i^{(n+1)}
\end{bmatrix}
& =\arg\dmax_{\beta,\beta_i} \left\{\exp \left[ -  h\left( \bu^{(n)}, \beta, \beta_i, i \in\Omega\setminus\{\bai\} \right) \right] \right\} \nonumber\\
& = \begin{bmatrix}
\ds\frac{\bu^{(n)^\dag}\bP_{\bv_0}^{\perp}\bx_{\bai}}{\bu^{(n)^\dag}\bP_{\bv_0}^{\perp}\bu^{(n)}}
\\
\ds \frac{\bu^{(n)^\dag}\bx_i}{\bu^{(n)^\dag}\bu^{(n)}}
\end{bmatrix}.
\label{eqn:iter2}
\end{align}
The iterative procedure starts by replacing $\bu^{(0)}$ with a normalized steering vector from the sidelobes.
Moreover, as for R-NCP-D, we stop alternating after a preassigned number of iterations.

Finally, replacing $\bM$ with the sample covariance matrix based upon the secondary data, we come up with an adaptive detector referred to in the following
as deterministic NCP detector (D-NCP-D).

\section{Performance Assessment}
\label{Sec:Assessment}
The aim of this section is to investigate the performance of the proposed algorithms 
in terms of probability of detection ($P_d$). To this end, we resort to standard Monte Carlo counting techniques by evaluating the thresholds
to ensure a preassigned $P_{fa}$ and the $P_d$ curves over $100/P_{fa}$ and $1000$ independent trials, respectively.
Data are generated according to the model defined in problem \eqref{eqn:HypProb}, where
\be
\bM = \sigma^2_n\bI_N +  p_c \bM_{c}.
\label{covMatrix}
\ee
In \eqref{covMatrix}, $\sigma^2_n\bI_N$ represents the thermal noise component with power $\sigma^2_n$ while  
$p_c \bM_{c}$ is the clutter component with
$p_c$ the clutter power and $\bM_c$ the structure of the clutter covariance matrix; the clutter-to-noise ratio (CNR) 
is thus given by
$\text{CNR}={p_c}/{\sigma^2_n}$.

In the following, we set $\sigma^2_n=1$,  $\text{CNR}=20$ dB, and $\bM_c(i,j)=\rho^{|i-j|}$ with $\rho=0.9$ (recall that $\bM_c(i,j)$ is the $(i,j)$th entry of $\bM_c$). 
Moreover, we suppose that the victim radar is equipped with 
a uniformly-spaced linear array (ULA) of $N$ identical and isotropic sensors with inter-element distance equal to $\lambda/2$, $\lambda$ being the wavelength corresponding to the carrier frequency of the NCP (modeled as a narrowband plane wave) impinging onto the antenna array.
The target response is computed according to the signal-to-clutter-plus-noise ratio (SCNR), whose
expression is
\be
\text{SCNR} = \left|\alpha \right|^2 \bv(0)^{\dag}\bM^{-1}\bv(0).
\label{SINR}
\ee
As for the NCP, we assume that, when it is present, it enters the antenna array response of the victim 
radar from the sidelobes with a power $p_j$ such that the jammer-to-noise ratio (JNR), defined as 
$\text{JNR}={p_j}/{\sigma^2_n}$,
is $30$ dB. In order to select the amount of primary data, we consider system parameter values of practical interest, namely we assume that
the radar system transmits a linear frequency modulated pulse of duration $3 \ \mu$s and bandwidth $5$ MHz.
Now, since the rate at which the range bins are generated is $2\cdot 10^{-7}$ s, then the uncompressed pulse covers $15$ range bins. 
Using $5$ additional guard cells, we set $H_1=H_2=10$. For the reader ease, all the simulation parameters are summarized in Table \ref{tab_parameters}.

Finally, for comparison purposes, we also report the $P_d$ curves of the subspace detector proposed in \cite{bandiera2007} and the adaptive matched filter 
(AMF) \cite{Kelly-Nitzberg}. As a matter of fact, the AMF raises from a suitable modification of the derivation contained 
in Subsection \ref{sec_mod_glrt_detector}, which consists in forcing the constraint $\bq^{\dag} \bM^{-1} \bv=0$ 
as shown in the appendix\footnote{Recall from Table \ref{tab_parameters} that the jammer is located at $35^\circ$ with respect to the array normal, 
leading to $\cos(\theta_{jt})=0.07$ for $N=8$ and $\cos(\theta_{jt})=0.03$ for $N=16$, where $\theta_{jt}$ is the angle between target and jammer
steering vectors in the whitened observation space.}. 
The curves for the likelihood ratio test
with known parameters, referred to as clairvoyant detector (CD), are also included since they represent an upper bound to the detection performance. 

In the next subsection, we focus on the stopping criterion and provide suitable numerical examples aimed at establishing a 
reasonable iteration number for both cyclic optimization procedures.

\subsection{Selection of the maximum number of iterations}
\begin{figure}
\begin{center}
\includegraphics[scale=0.36]{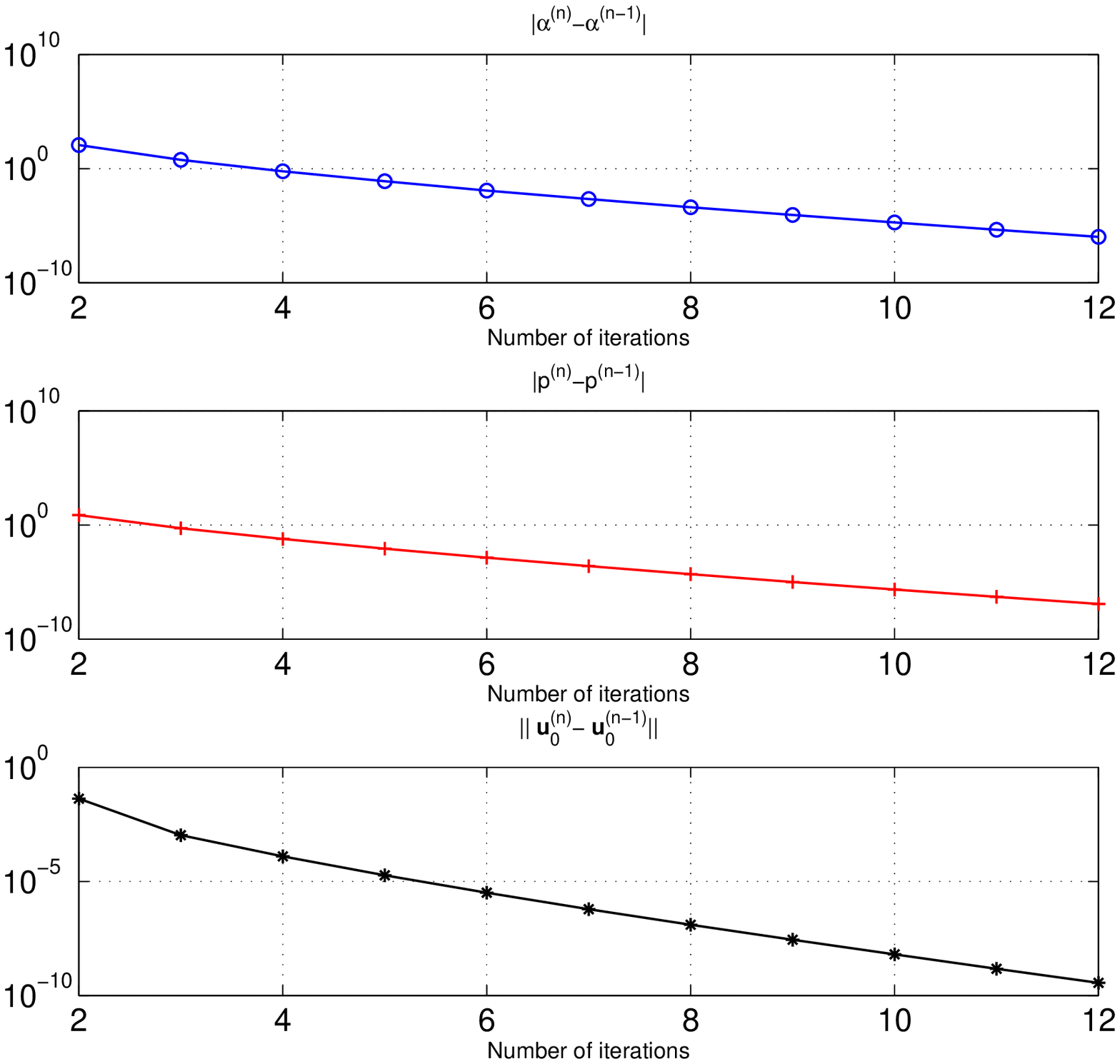}
\caption{Euclidean norm (or modulus) of the difference between the estimates at the $n$th iteration and 
those at the $(n-1)$th iteration versus the number of iterations for cyclic optimization 
of Subsection \ref{sec_glrt_detector}.}
\label{fig:convergenceP1}
\end{center}
\end{figure}

\begin{figure}
\begin{center}
\includegraphics[scale=0.36]{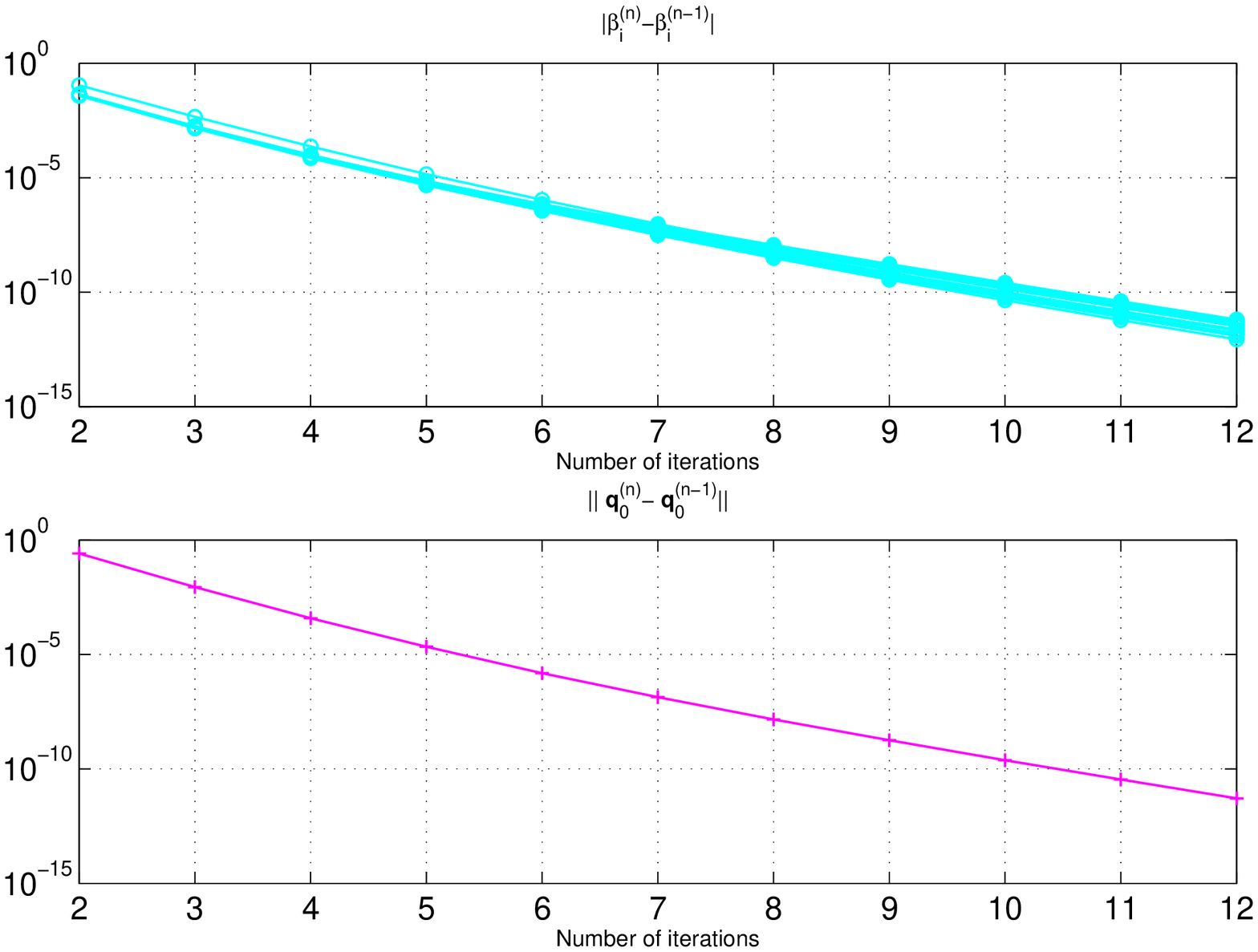}
\caption{Euclidean norm (or modulus) of the difference between the estimates at the $n$th iteration and 
those at the $(n-1)$th iteration versus the number of iterations for cyclic optimization 
of Subsection \ref{sec_mod_glrt_detector}.}
\label{fig:convergenceP7}
\end{center}
\end{figure}

The detection performance, assessed in Subsection \ref{subsec:detectionPerformance}, are obtained
using a preassigned number of iterations. 
Now, in order to select this value, in Figures \ref{fig:convergenceP1} and \ref{fig:convergenceP7}, we show the average norm
of the difference between the estimates at the $n$th iteration and their respective values at the $(n-1)$th iteration. The averages are evaluated over $1000$
independent Monte Carlo trials assuming $N=8$, $K=12$, and SCNR$=20$ dB. In both cases we model $\bq$ as a narrowband plane wave impinging onto the antenna array from a direction whose azimuth, generated at random at each Monte Carlo trial, is uniformly distributed outside the antenna mainlobe. Inspection of the figures highlights that $10$ iterations for each procedure
ensure a variation of the estimated quantities less than or equal to $10^{-5}$ and also represent a good compromise from the computational point of view.
For this reason, in what follows we adopt this number for the computation of the $P_d$ curves.

\subsection{Detection Performance}
\label{subsec:detectionPerformance}
In this subsection, we investigate the behavior of the proposed architectures in terms of $P_d$ versus the SCNR for 
two different scenarios, which are complementary. 
Specifically, the former assumes a jammer illuminating the victim radar from its 
sidelobes, whereas the latter consider a surveillance area 
free of intentional interferers. It is important to underline that the second case does not correspond to the design assumptions.

\begin{figure}
\begin{center}
\includegraphics[scale=0.35]{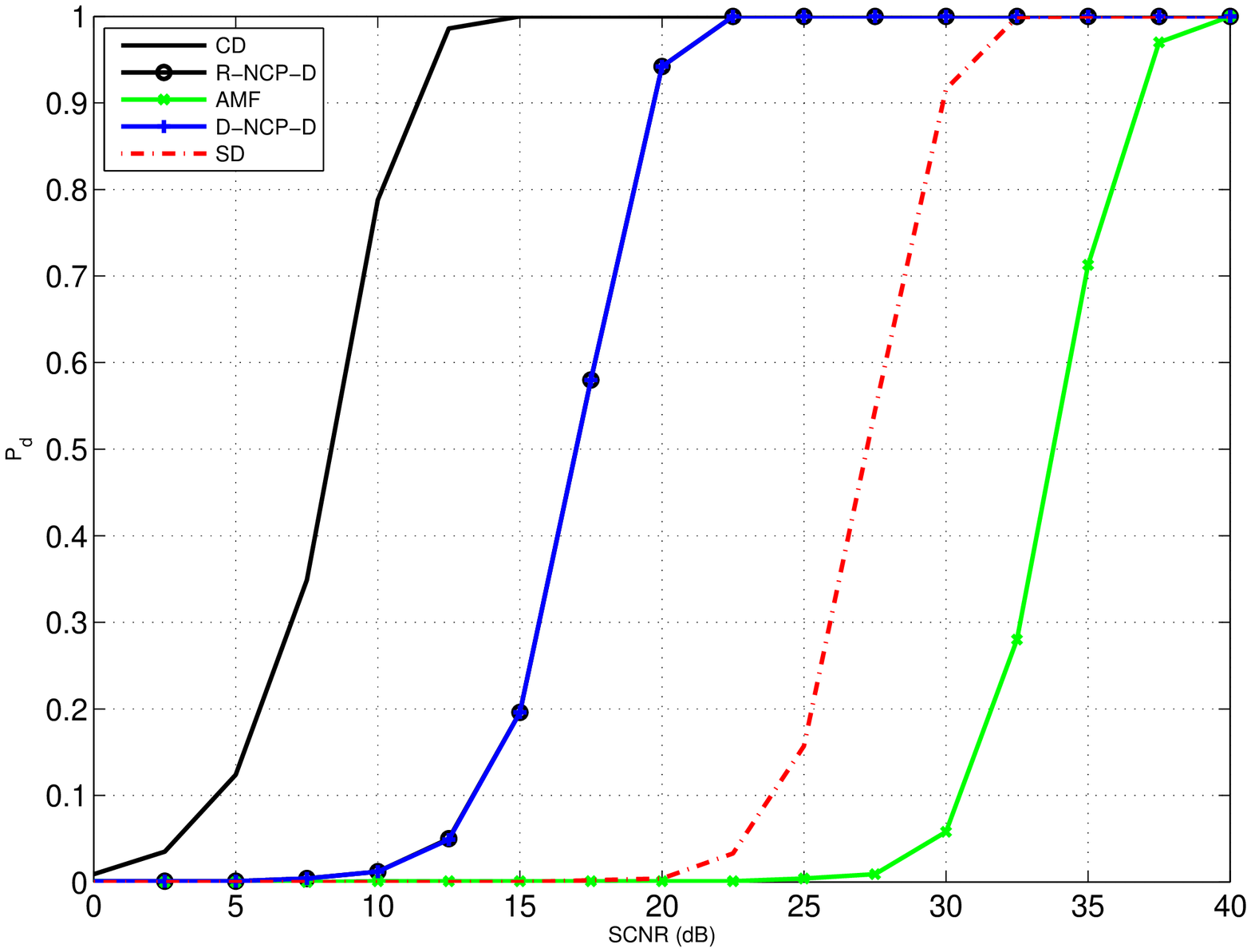}
\caption{$P_d$ versus SCNR for the CD, R-NCP-D, AMF, D-NCP-D, and the SD assuming $N=8$, $K=12$, and a jammer at $35^\circ$.}
\label{fig:N8K12_NCP}
\end{center}
\end{figure}

\begin{figure}
\begin{center}
\includegraphics[scale=0.35]{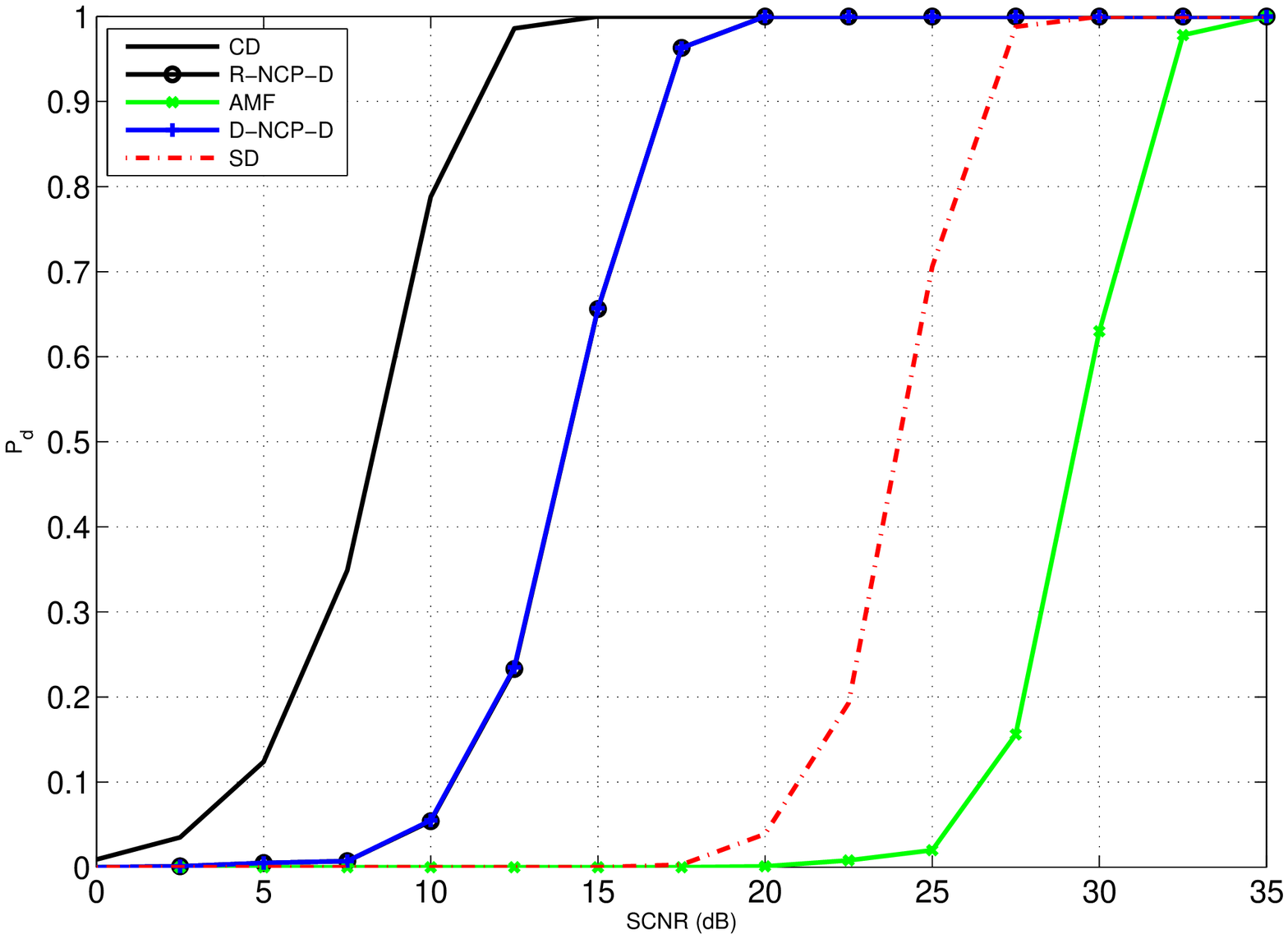}
\caption{$P_d$ versus SCNR for the CD, R-NCP-D, AMF, D-NCP-D, and the SD assuming $N=8$, $K=16$, and a jammer at $35^\circ$.}
\label{fig:N8K16_NCP}
\end{center}
\end{figure}

\begin{figure}
\begin{center}
\includegraphics[scale=0.35]{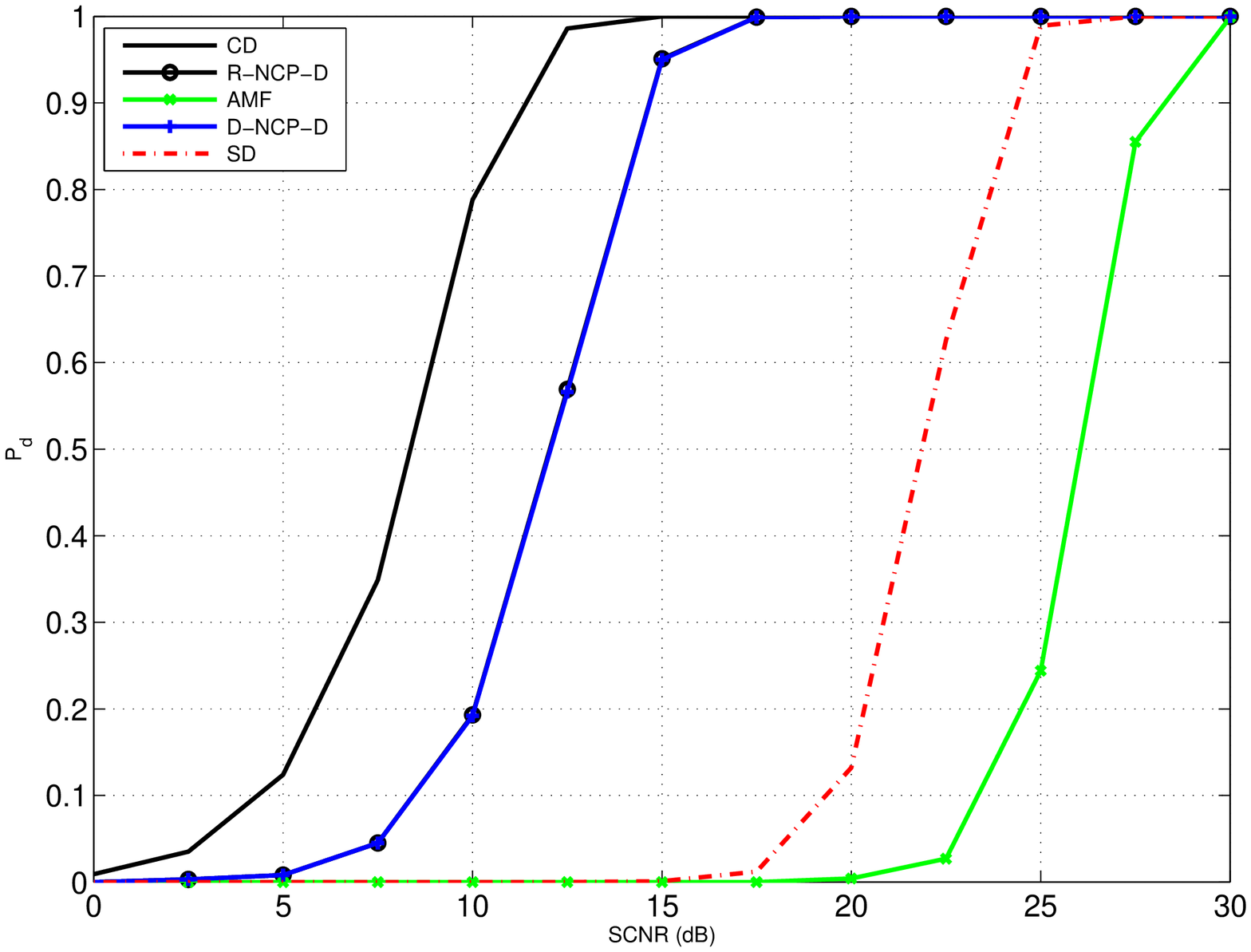}
\caption{$P_d$ versus SCNR for the CD, R-NCP-D, AMF, D-NCP-D, and the SD assuming $N=8$, $K=24$, and a jammer at $35^\circ$.}
\label{fig:N8K24_NCP}
\end{center}
\end{figure}

Figures \ref{fig:N8K12_NCP}-\ref{fig:N8K24_NCP} refer to the first scenario assuming $N=8$ and different values for $K$. The common denominator of 
these figures
is that R-NCP-D and D-NCP-D share the same performance, since the respective curves are overlapped, and 
outperform the remaining architectures except for the CD (as expected). 
The gain of the R-NCP-D and D-NCP-D over the SD is about $10$ dB at $P_d=0.9$. On the other hand, the AMF has the worst performance 
with a loss at $P_d=0.9$ ranging
from about $7$ dB for $K=12$ to about $4$ dB for $K=24$ with respect to the SD. Comparison of the figures also points out that
the $P_d$ curves move towards left as $K$ increases, which means that the $P_d$ is an increasing function of $K$ given the SCNR. 
Finally, the loss at $P_d=0.9$ of the R-NCP-D/D-NCP-D with respect to the CD halves when $K$ goes from $K=12$, which corresponds to a loss of about $8$ dB, 
to $K=24$, which results in a loss of about $4$ dB.

\begin{figure}
\begin{center}
\includegraphics[scale=0.35]{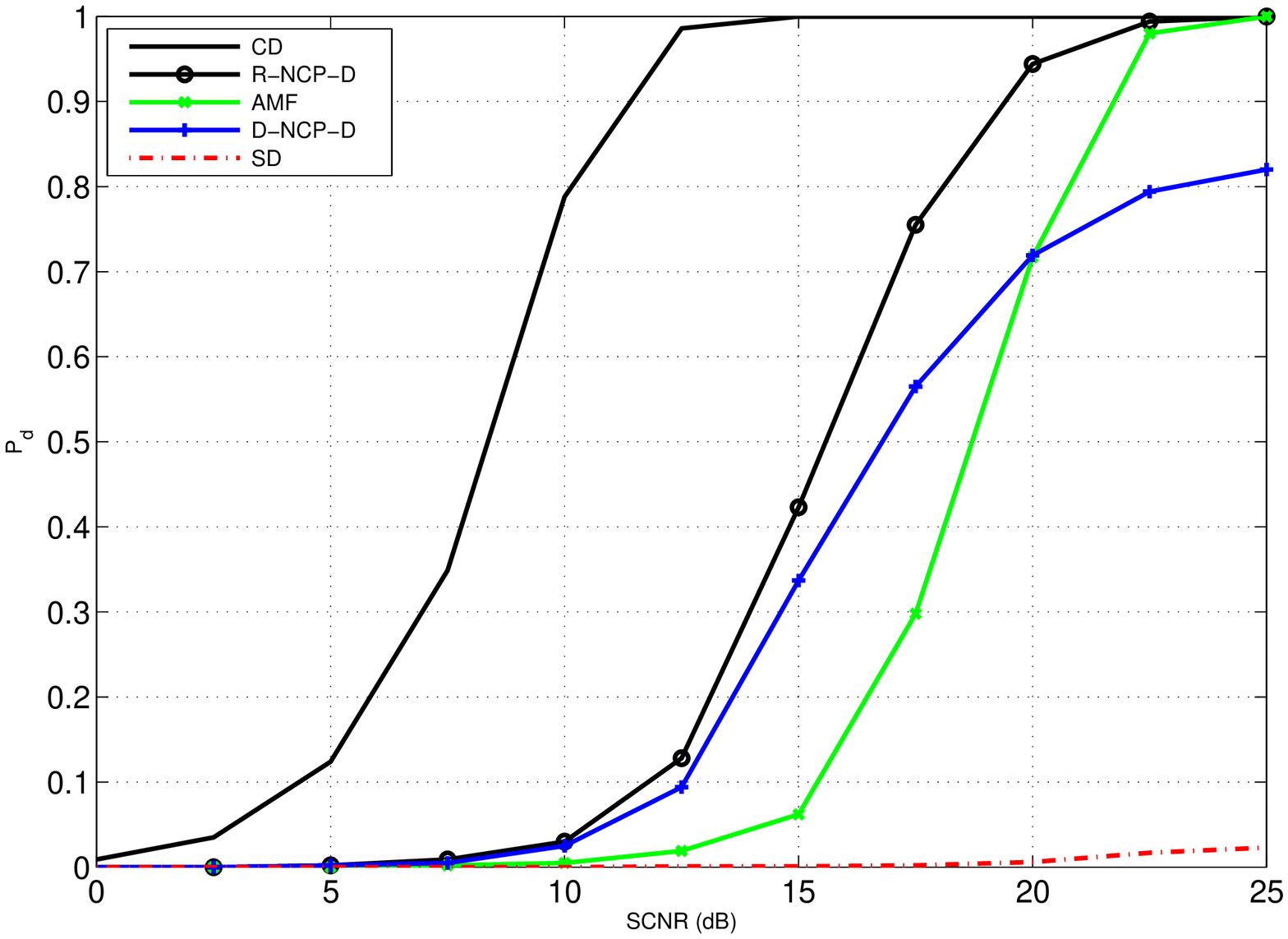}
\caption{$P_d$ versus SCNR for the CD, R-NCP-D, AMF, D-NCP-D, and the SD assuming $N=8$, $K=12$, and no jammers.}
\label{fig:N8K12_NO_NCP}
\end{center}
\end{figure}

\begin{figure}
\begin{center}
\includegraphics[scale=0.35]{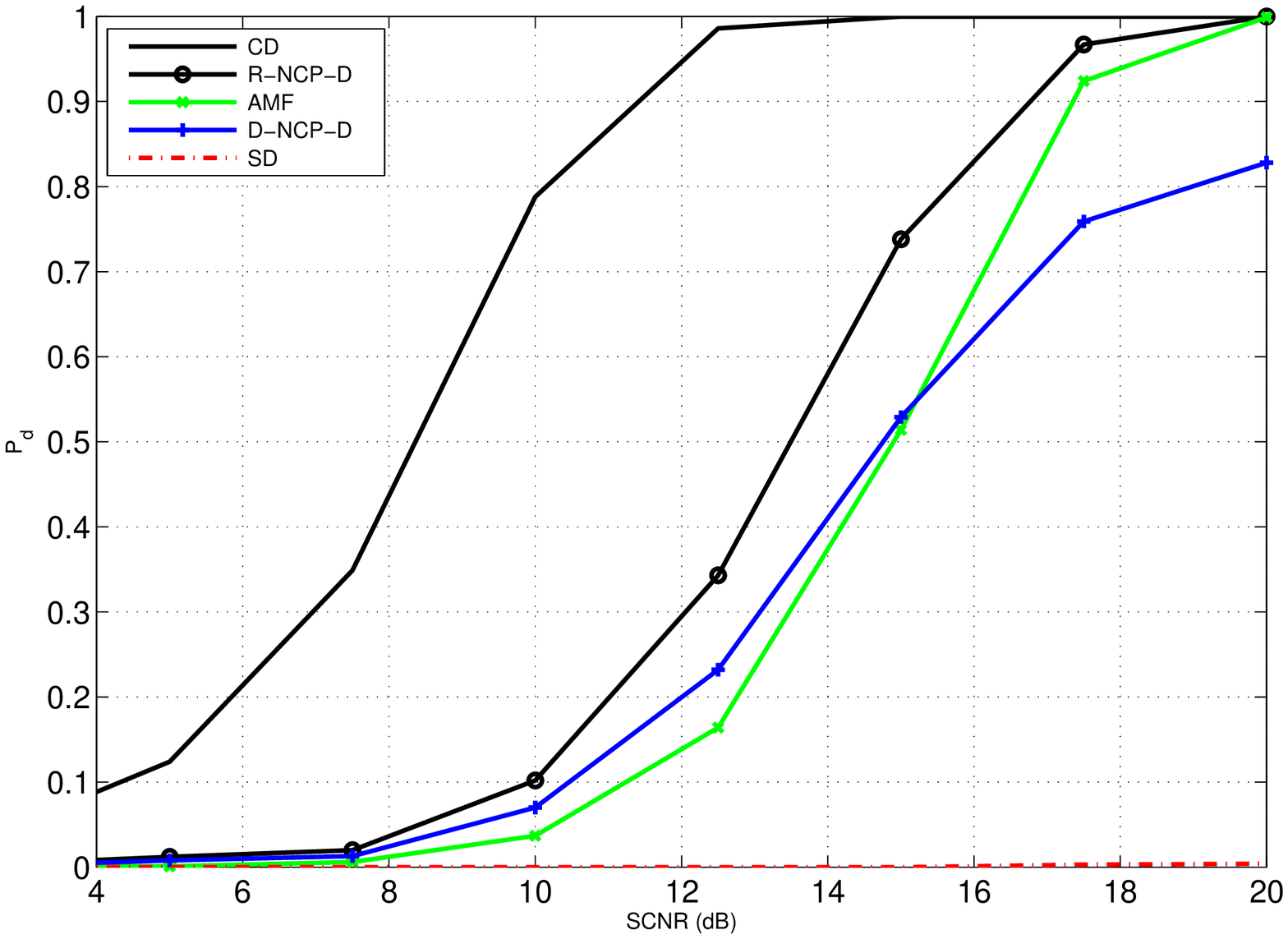}
\caption{$P_d$ versus SCNR for the CD, R-NCP-D, AMF, D-NCP-D, and the SD assuming $N=8$, $K=16$, and no jammers.}
\label{fig:N8K16_NO_NCP}
\end{center}
\end{figure}

\begin{figure}
\begin{center}
\includegraphics[scale=0.35]{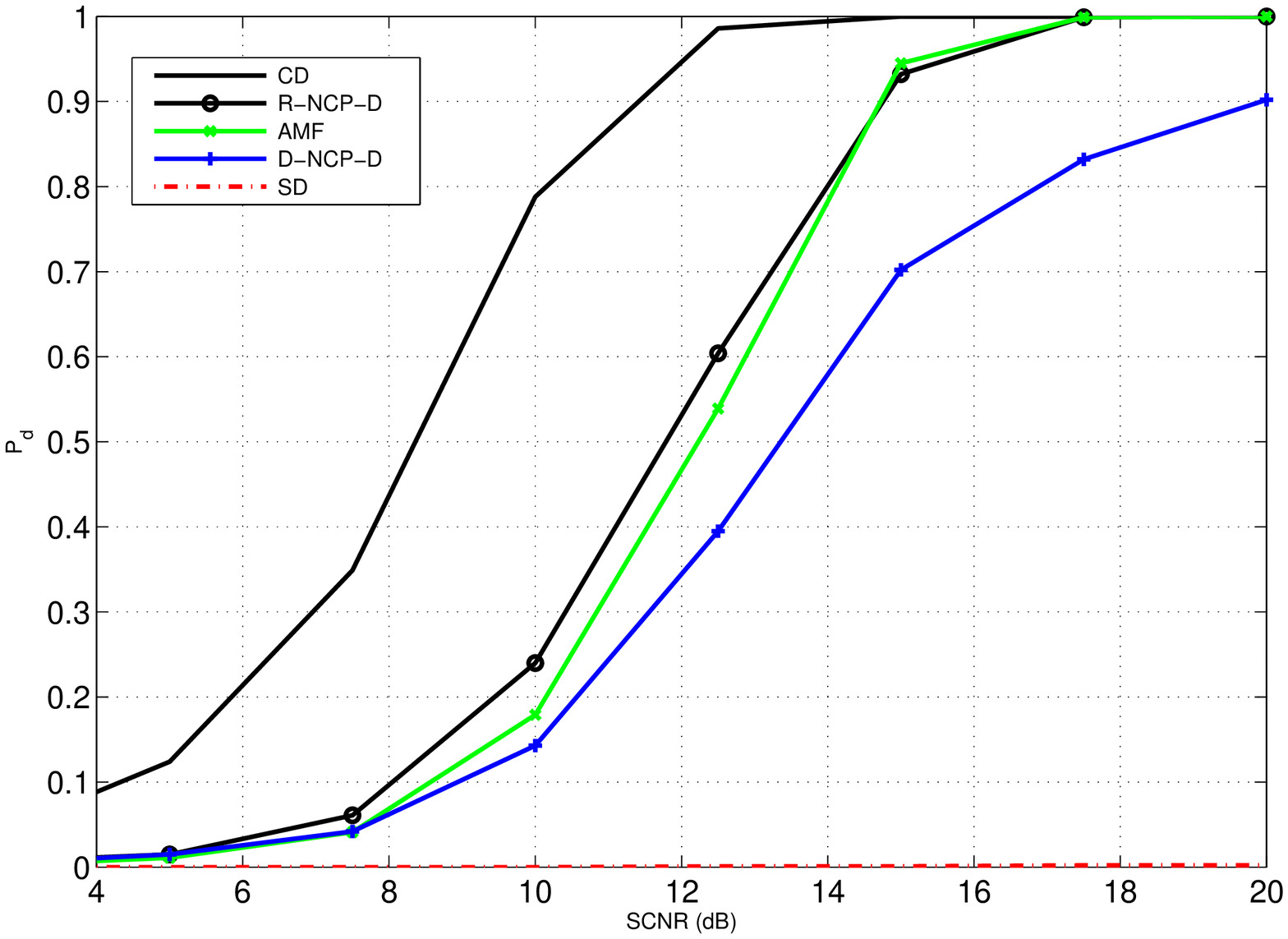}
\caption{$P_d$ versus SCNR for the CD, R-NCP-D, AMF, D-NCP-D, and the SD assuming $N=8$, $K=24$, and no jammers.}
\label{fig:N8K24_NO_NCP}
\end{center}
\end{figure}

The second scenario is accounted for in Figures \ref{fig:N8K12_NO_NCP}-\ref{fig:N8K24_NO_NCP}, which assume the same parameter setting as
previous figures except for the presence of the jammer. In this case, the curves of R-NCP-D and D-NCP-D are no longer overlapped and, more important,
the latter does not achieve $P_d=1$ at least for the considered parameter values. The R-NCP-D continues to provide satisfactory performance
outperforming the other counterparts. In fact, the comparison with previous figures highlights that the performance of R-NCP-D keeps more or less unaltered.
On the other hand, for the considered parameter values (and the considered scenario), 
the SD is completely useless since the resulting $P_d$ values are close to zero, while
the AMF improves significantly its performance as $K$ increases ensuring about the same $P_d$ values of the R-NCP-D for $K=3N$.

\begin{figure}
\begin{center}
\includegraphics[scale=0.35]{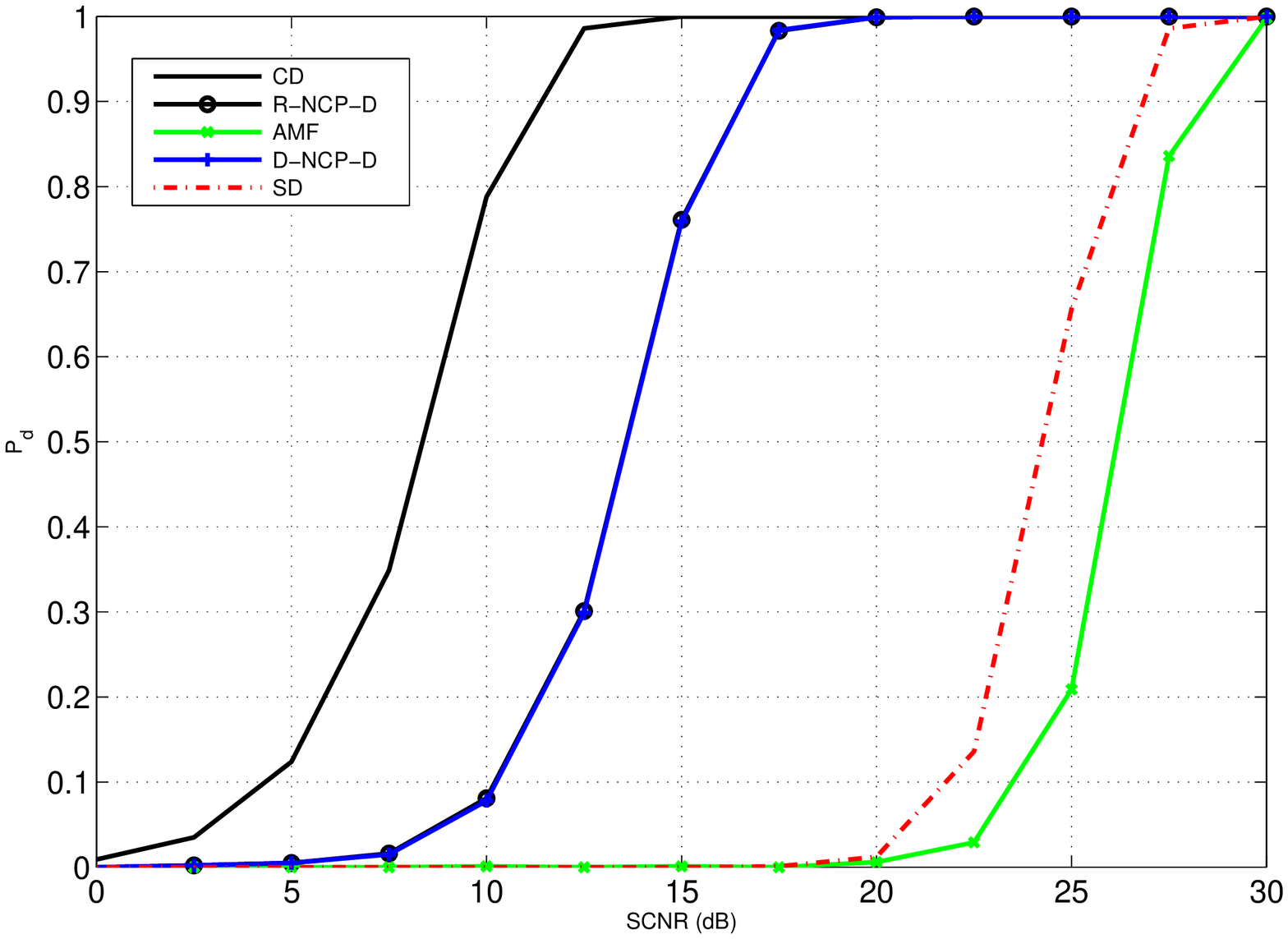}
\caption{$P_d$ versus SCNR for the CD, R-NCP-D, AMF, D-NCP-D, and the SD assuming $N=16$, $K=32$, and a jammer at $35^\circ$.}
\label{fig:N16K32_NCP}
\end{center}
\end{figure}

Finally, the comparison between Figure 
\ref{fig:N16K32_NCP} and Figure \ref{fig:N8K16_NCP} allows to appreciate the performance 
variations due to both $N$ and $K$ when their ratio is constant. In fact, Figure 
\ref{fig:N16K32_NCP} assumes $N=16$ and $K=32$, namely twice the analogous values of Figure \ref{fig:N8K16_NCP}.
It can be noted that R-NCP-D, D-NCP-D, and AMF provides higher $P_d$ values than those of Figure \ref{fig:N8K16_NCP}, whereas the
performance of SD seems insensitive to the considered parameter change.

Summarizing, the above analysis singles out the R-NCP-D as an effective mean to face attacks of smart NLJs which 
transmit noise-like signals to cover the skin echoes from the platform under protection. As a matter of fact,
the R-NCP-D outperforms all the other competitors either in the presence or absence of a noise cover pulse.

\section{Conclusions}
\label{Sec:Conclusions}
In this paper, two new detection architectures to account for possible NCP attacks have been proposed and assessed. Specifically, 
the first approach consists in modeling the NLJ contribution as a covariance component, while the second solution considers 
the realizations of the NLJ and handles them as deterministic signals. Then, ad hoc modifications of the 
GLRT have been devised for both scenarios where the unknown parameters are computed through alternating estimation procedures.
The behavior of the two architectures has been first
investigated resorting to simulated data adhering the design assumptions of the first approach and, then, they have been tested on 
data where the NCP is turned off.  
The analysis has singled out the R-NCP-D obtained with the first approach 
as the recommended solution for adaptive detection in the presence of clutter and NCP,
since it can guarantee satisfactory performance in both the considered situations.

Finally, future research tracks might encompass the application of the herein presented approach 
to the case of range-spread targets possibly embedded in non-Gaussian clutter.

\begin{appendices}
\section{Alternative derivation of the AMF}
\label{Appendix_sec_constrained}

In this Appendix, we modify the derivation contained in Subsection \ref{sec_mod_glrt_detector} by imposing the constraint $\bu^{\dag}\bv_0=0$, 
namely the target steering vector and the NCP signature are orthogonal in the whitened space.
For the reader convenience, let us recall that $\bu=\bM^{-1/2}\bq$, $\bS_1=\left[\bx_{\bai}\bx_{\bai}^{\dag}+\sum_{i \in\Omega\setminus\{\bai\}}\bx_{i}\bx_{i}^{\dag}\right]$,
$\bv_0=\bM^{-1/2}\bv$, $\bx_i=\bM^{-1/2}\bz_i$, $\bx_{\bai}=\bM^{-1/2}\bz_{\bai}$.

Now, under $\cH_0$, the maximization with respect to $\beta$ and $\beta_i$ of $w$ (given by \eqref{eq_w}) 
leads to \eqref{eq_minbeta} and \eqref{eq_minbetai}, respectively. Thus, assuming the orthogonality constraint, 
it is possible to reformulate the maximization of $w$ with respect to $\bu$ as 
\be
\max\limits_{\bu:\bu^{\dag}\bv_0=0} \exp \left\{  \frac{\bu^{\dag}\bS_1\bu}{\bu^{\dag}\bu} - \tr[\bS_1]
   \right\}.
\label{eq_S1}
\ee
Let $\bU\in\C^{N\times N-1}$ be a slice of unitary matrix, namely $\bU^{\dag}\bU=\bI_{N-1}$, with columns 
forming a basis for the orthogonal complement of the subspace spanned by $\bv_0$. Given the orthogonality constraint, it follows that
$\bu$ can be expressed in terms of a linear combination of the columns of $\bU$, i.e., $\bu=\bU\mathbf{\bgamma}$ with $\bgamma\in\C^{(N-1)\times1}$
the coordinate vector. Then, maximization \eqref{eq_S1} can be recast as
\begin{align}
& \max\limits_{\bu:\bu^{\dag}\bv_0=0} \exp \left\{  \frac{\bu^{\dag}\bS_1\bu}{\bu^{\dag}\bu} - \tr[\bS_1]
\right\} \nonumber\\ 
& =
\exp \left\{ 
\max_{\bgamma} \frac{\bgamma^{\dag}\bU^{\dag}\bS_1\bU\bgamma}{\bgamma^{\dag}\bgamma}
- \tr[\bS_1]
\right\}
\nonumber \\ &=
\exp \left\{ 
\lambda_1\left( \bU^{\dag}\bS_1 \bU \right)
- \tr[\bS_1]
\right\},
\label{eq_maxnu}
\end{align}
where the last equality is due to the Rayleigh Ritz theorem \cite{HornJohnson}.
Thus, the solution to the constrained optimization problem, under $\cH_0$, is
\begin{align}
& \max\limits_{\beta,\beta_i, i \in\Omega\setminus\{\bai\},
\bq}\cL_0( \beta,\beta_i,
i \in\Omega\setminus\{\bai\},
\bM,\bq) \nonumber \\ 
&= 
\frac{1}{\left[\pi^N \det \left(\bM \right) \right]^{H}}
\exp \left\{  \lambda_1\left( \bU^{\dag}\bS_1 \bU \right) - \tr[\bS_1]
\right\}.
\label{eq_solH0ort}
\end{align}
On the other hand, under $\cH_1$, we can start from
the partially-compressed likelihood with respect to $\alpha$, given by 
(\ref{eq:compressed_L1_alpha}).
In fact, it is possible to show
that, after optimization of the likelihood function with respect to 
$\alpha$, 
$\beta$, and $\beta_i$, 
we obtain 
\begin{align}
\label{maxonecol}
& \max\limits_{\alpha,\beta,\beta_i, i \in\Omega\setminus\{\bai\},
\bq}\cL_1( \alpha,\beta,\beta_i,
i \in\Omega\setminus\{\bai\},
\bM,\bq)  \nonumber \\ 
& = \frac{1}{\left[\pi^N \det \left(\bM \right) \right]^{H}}
\exp \left\{
\frac{\bu^{\dag}\bP_{v_0}^{\perp} \bx_{\bai}\bx_{\bai}^{\dag}\bP_{v_0}^{\perp} \bu}{\bu^{\dag}\bP_{v_0}^{\perp} \bu}  \right. \nonumber\\
& \left. + \sum\limits_{i \in\Omega\setminus\{\bai\}} \frac{\bx_i^{\dag}\bu\bu^{\dag}\bx_i}{\bu^{\dag}\bu} - \left( \bx_{\bai}^{\dag}\bP_{v_0}^{\perp} \bx_{\bai} +  \sum\limits_{i \in\Omega\setminus\{\bai\}}\bx_i^{\dag}\bx_i \right)\right\} 
\nonumber \\
&= \frac{1}{\left[\pi^N \det \left(\bM \right) \right]^{H}}
\exp \!\! \left\{
\frac{\bu^{\dag}\bS_1 \bu}{\bu^{\dag} \bu} - \left( \bx_{\bai}^{\dag}\bP_{v_0}^{\perp} \bx_{\bai} +  \!\!\!\! \sum\limits_{i \in\Omega\setminus\{\bai\}} \!\!\! \bx_i^{\dag}\bx_i \right)\!\!\! \right\} 
\end{align}
where the last equality comes from the orthogonality condition between $u$ and $v_0$.
Thus, it follows that
\begin{align}
& \dmax_{\bf u:u^\dag v_0=0}
\max\limits_{\alpha,\beta,\beta_i, i \in\Omega\setminus\{\bai\},
\bq}\cL_1( \alpha,\beta,\beta_i,
i \in\Omega\setminus\{\bai\},
\bM,\bq) 
\nonumber \\
&= \frac{1}{\left[\pi^N \det \left(\bM \right) \right]^{H}}
\nonumber \\ 
& \times
\exp \left\{
\lambda_1 ( \bU^\dag\bS_1 \bU) - \left( \bx_{\bai}^{\dag}\bP_{v_0}^{\perp} \bx_{\bai} +  \sum\limits_{i \in\Omega\setminus\{\bai\}}\bx_i^{\dag}\bx_i \right)\right\} \label{eq_solH1ort}
\end{align}
where, again, the last equality comes from \cite{HornJohnson}.
The final expression for the constrained GLRT-based architecture is obtained combining \eqref{eq_solH0ort} and \eqref{eq_solH1ort}. Precisely, taking the log-likelihoods, the ``constrained test \eqref{eq_glrtbis}'' is statistically equivalent to
\be
\bx_{\bai}^{\dag}\bP_{v_0}\bx_{\bai} \test \eta,
\label{eq_glrtbis2}
\ee
whose decision statistic, after replacing $\bM$ with the sample covariance matrix based on secondary data, coincides with that of the AMF.



\end{appendices}

\begin{table}[H]
\centering
\caption{Simulation Parameters Setup}
\begin{tabular}{|l|l|l|}
\hline
\textbf{Symbol} & \textbf{Value} & \textbf{Description} \\ \hline \hline
$N$             & 8,16              & Number of antennas                                                          \\ \hline
$H_1$       & 10             & Number of contaminated cells                                \\ 
		&			& 	on the left of the CUT  \\ \hline
$H_2$       & 10             & Number of contaminated cells                                 \\ 
		&			& 	on the right of the CUT  \\ \hline
CNR             & 20 dB          & Clutter-to-Noise Ratio                                                      \\ \hline
JNR             & 30 dB          & Jammer-to-Noise Ratio                                                       \\ \hline
$\theta_j$      & 35$^{\circ}$   & Angle Of Arrival (AOA) of the NCP jammer                                                       \\ \hline
$P_{fa}$        & 10$^{-4}$      & False Alarm probability                                                      \\ \hline
$N_{MC}$        & 10$^3$         & Number of independent trials \\ 
		&			& 	to evaluate the probability of detection $P_d$   \\ \hline
\end{tabular}
\label{tab_parameters}
\end{table}
\end{document}